\documentclass[12pt]{article}
\usepackage{amsmath,amsthm,amsfonts,amssymb, color}
\usepackage{algorithm}
\usepackage{algorithmicx}
\usepackage{algpseudocode}
\usepackage{booktabs}
\usepackage{graphicx} % For resizing the table
\usepackage{adjustbox} % For advanced resizing and alignment
\usepackage{tcolorbox}
\usepackage{hyperref}
\usepackage{mathtools}
\usepackage{float}
\usepackage{makecell}
\usepackage{xcolor}
\usepackage{pgfplots}
\usepackage{pgfplotstable}
\pgfplotsset{compat=1.17}
\usepackage{pgfplots}
\usepackage{pgfplotstable}

\usepackage{caption}
\usepackage{subcaption}
\usepackage{pdfpages}
\usepackage{tikz}% for plot Interval
\usepackage{graphicx}
\usepackage{overpic}
\usepackage{tikz-cd}  %used for the diagram
\usepackage{authblk}

\newtheorem{theorem}{Theorem}[section]

\newtheorem{definition}[theorem]{Definition}
\newtheorem{remark}[theorem]{Remark}

\begin{document}
%\tableofcontents

\title{ProT-GFDM: A Generative Fractional Diffusion Model for Protein Generation}

\author[1]{\small Xiao Liang}
\author[2]{\small Wentao Ma}
\author[3,4]{\small Eric Paquet}
\author[4]{\small Herna Lydia Viktor}
\author[1]{\small Wojtek Michalowski}

\affil[1]{\footnotesize Telfer School of Management, University of Ottawa, Ottawa, ON, K1N 6N5, Canada}
\affil[2]{\footnotesize Department of Computer Science, University of Toronto, Toronto, ON, M5S 2E4, Canada}
\affil[3]{\footnotesize National Research Council, 1200 Montreal Road, Ottawa, ON, K1A 0R6, Canada}
\affil[4]{\footnotesize School of Electrical Engineering and Computer Science, University of Ottawa, Ottawa, ON, K1N 6N5, Canada}

%\author{Xiao Liang\footnote{University of Ottawa, Department of Mathematics and Statistics, 150 Louis Pasteur Private, Ottawa, Ontario, K1G 0P8, Canada}, 
%Wentao Ma\footnote{University of Toronto, Department of Computer Science, 40 St George St, Toronto, ON, M5S 2E4, Canada},
%Eric Paquet\footnote{National Research Council, 1200 Montreal Road, Ottawa, ON, K1A 0R6, Canada, School of Electrical Engineering and Computer Science, University of Ottawa, ON, K1N 6N5, Canada},
%Herna Lydia Viktor\footnote{School of Electrical Engineering and Computer Science, University of Ottawa, ON, K1N 6N5, Canada},
%Wojtek Michalowski\footnote{Telfer School of Management, University of Ottawa, ON, K1N 6N5, Canada}.
%}

\date{\today}
\maketitle

%% Abstract
\begin{abstract}
This work introduces the generative fractional diffusion model for protein generation (ProT-GFDM), a novel generative framework that employs fractional stochastic dynamics for protein backbone structure modeling. This approach builds on the continuous-time score-based generative diffusion modeling paradigm, where data are progressively transformed into noise via a stochastic differential equation and reversed to generate structured samples. Unlike classical methods that rely on standard Brownian motion, ProT-GFDM employs a fractional stochastic process with superdiffusive properties to improve the capture of long-range dependencies in protein structures. Trained on protein fragments from the Protein Data Bank, ProT-GFDM outperforms conventional score-based models, achieving a 7.19\% increase in density, a 5.66\% improvement in coverage, and a 1.01\% reduction in the Fr\'{e}chet inception distance. By integrating fractional dynamics with computationally efficient sampling, the proposed framework advances generative modeling for structured biological data, with implications for protein design and computational drug discovery. 
%Code available at: \url{https://github.com/iamtonymwt/ProGFDM}.  
\end{abstract}



%% Keywords




\vspace{1mm}



\section{Introduction}

Proteins are fundamental macromolecules for diverse biological functions. Their unique amino acid sequences have three-dimensional (3D) structures, which determine cellular roles, specificity, and stability \cite{Morris2022, Luo2023}. {\em De novo} \cite{PoSsu2016} protein generation aims to produce new protein sequences that fold into stable and functional structures, creating novel biomolecules for applications in medicine, biotechnology, and synthetic biology. Traditionally, protein design follows a two-step process: determining a backbone structure that meets the desired structural and biochemical properties and designing an amino acid sequence to fit that backbone. Often reliant on template-based fragment sampling and expert-defined topologies, this approach has limitations, including that the backbone may not be optimally designable or the sequence may fail to conform to the intended structure. These problems can lead to a relatively low success rate in computational protein design \cite{Lai2024}. During the past decades, the exponential growth of protein sequence data has driven the development of myriad computational methods for de novo protein design \cite{Pan2021} and protein function prediction \cite{Lai2024}. A systematic review and comparison of methods is provided in \cite{Lin2024}.

%Existing computational modeling techniques for protein design often rely on time-intensive simulations, heuristic-driven algorithms, and extensive hand-tuning. 
\medskip

Recent progress in artificial intelligence (AI) models has substantially changed the field of protein design. Traditionally, generative models, such as generative adversarial networks (GANs) \cite{GAN1-2014}, variational autoencoders (VAEs) \cite{VAE-2013}, and flow-based models \cite{Flow-2015} have been widely applied to protein generation, enabling the design of novel proteins with the desired properties and facilitating the exploration of vast combinatorial spaces. For example, ProteinVAE \cite{VAE1-2022} applies ProtBERT \cite{VAE2-2021} to convert raw protein sequences into latent representations, using an encoder–decoder framework enhanced with positionwise multihead self-attention to capture long-range sequence dependencies. In contrast, ProT-VAE \cite{VAE3-2023} employs a different pretrained language model, ProtT5NV, and incorporates an internal, family-specific encoder–decoder layer to learn parameters tailored to individual protein families. Conversely, ProteinGAN \cite{Prot-GAN-2021} employs a GAN architecture to generate protein sequences, and its capabilities are illustrated through the example of malate dehydrogenase, demonstrating its potential to produce fully functional enzymes. 

However, Strokach et al. listed the disadvantages of generative models for protein design \cite{Strokach2022}. For instance, GANs and VAEs each have their drawbacks. For example, GANs can experience unstable training processes compared to other model types and often generate examples with low diversity due to mode collapse. Additionally, GANs lack mechanisms for mapping existing data to latent spaces or for calculating log-likelihoods. In contrast, VAEs are typically less effective at modeling data with a fixed dimensionality and often generate lower-resolution samples than GANs. Other deep learning approaches have also played a significant role in advancing protein design (e.g., \cite{Wang-2018, Qi-2020, Anishchenko-2021, Dauparas-2022, Ferruz-2022}).

\medskip

Probabilistic diffusion models have recently demonstrated significant potential in bioinformatics, particularly in such applications as protein generation. These models learn to approximate complex data distributions by sequentially corrupting and denoising data samples, enabling the synthesis of high-quality, realistic output \cite{Song-ICLR-2021}. {\bf Denoising diffusion probabilistic models} (DDPMs) \cite{Jonathan-2020} define a discrete-time Markov chain that progressively adds Gaussian noise to data, transforming data into a nearly isotropic Gaussian distribution. Formally, for each training sample $\mathbf{x}_0 \sim p_{\text{data}}(\mathbf{x})$, the forward process follows a predefined variance schedule $0 < \beta_1 < \beta_2, \dots, \beta_T < 1$ and constructs a sequence $\{ \mathbf{x}_0, \mathbf{x}_1, \dots, \mathbf{x}_T \}$, where the transition at each step is given by the following:
$$
p(\mathbf{x}_t | \mathbf{x}_{t-1}) = \mathcal{N}(\mathbf{x}_t; \sqrt{1 - \beta_t} \mathbf{x}_{t-1}, \beta_t \mathbf{I}).
$$
Expanding this recurrence yields direct mapping from $\mathbf{x}_0$ to $\mathbf{x}_t$:
\begin{equation}
\label{discrete_DDPM}
\mathbf{x}_t = \sqrt{1 - \beta_t} \mathbf{x}_{t - 1} + \sqrt{\beta_t} \boldsymbol{\varepsilon}_{t-1}, \quad t = 1, \dots, T,
\end{equation}
where $\boldsymbol{\varepsilon}_{t} \sim \mathcal{N}(0, \bold{I})$. Applying the \textbf{Markov property}, we can marginalize over all intermediate steps and derive the closed-form distribution of $\mathbf{x}_t$, given the original data sample:
$$
p(\mathbf{x}_t | \mathbf{x}_0) = \mathcal{N}(\mathbf{x}_t; \sqrt{\alpha_t} \mathbf{x}_0, (1 - \alpha_t) \mathbf{I}),
$$
where $\alpha_t = \prod_{s=1}^{t} (1 - \beta_s)$. The noise schedule is predefined such that $\alpha_T \to 0$, ensuring that $\mathbf{x}_T$ asymptotically approaches a standard Gaussian distribution (i.e., $p(\mathbf{x}_T) \approx \mathcal{N}(0, \mathbf{I})$). The \textbf{reverse process}, which transforms Gaussian noise back into structured data, is modeled as another Markov chain, parameterized as follows:
$$
p(\mathbf{x}_{t-1} | \mathbf{x}_t) = \mathcal{N}(\mathbf{x}_{t-1}; \mu_{\theta}(\mathbf{x}_t, t), \boldsymbol{\Sigma}(t)),
$$
where the mean function $\mu_{\theta}(\mathbf{x}_t, t)$ is learned using a neural network $\text{NN}_{\theta}(\mathbf{x}_t, t)$. Once the model is trained, new samples can be generated by first drawing $
\mathbf{x}_T \sim \mathcal{N}(0, \mathbf{I})
$, and iteratively updating:
$$
\mathbf{x}_{t-1} = \frac{1}{\sqrt{1 - \beta_t}} \left( \mathbf{x}_t + \beta_t \, \text{NN}_{\theta^{\ast}}(\mathbf{x}_t, t) \right) + \sqrt{\beta_t} \mathbf{z}_t, \quad t = T, T-1, \dots, 1,
$$
where $\theta^{\ast}$ denotes the optimized parameters of the trained network $\text{NN}_{\theta}$. The DDPM effectively recovers structured data from noise via this iterative denoising process. An alternative approach to generative modeling is \textbf{score-matching Langevin dynamics (SMLD)} \cite{YS-SE-2019-dd}, a method initially developed in physics to describe the stochastic motion of particles under deterministic and random forces. Langevin dynamics facilitates sampling from a target distribution $p(\mathbf{x})$ using an iterative update rule:
\begin{equation}
\label{SMLD}
\mathbf{x}_{t} = \mathbf{x}_{t-1} + \tau \nabla_{\mathbf{x}} \log p(\mathbf{x}_{t-1}) + \sqrt{2 \tau} \mathbf{z}, \quad \mathbf{z} \sim \mathcal{N}(0, \mathbf{I}),
\end{equation}
where $\tau$ denotes the step size, and $\mathbf{x}_0$ is initialized from white noise. The term $\nabla_{\mathbf{x}} \log p(\mathbf{x})$ represents the \textbf{score function}, also known as \textbf{Stein’s score function}, which determines the optimal update direction for the sample. The score function $\nabla_{\mathbf{x}} \log p(\mathbf{x})$ must be learned using neural networks to apply Langevin dynamics for generative modeling. Once trained, samples can be generated by iteratively refining noisy input via Langevin dynamics, guiding them toward the target data distribution. Yang et al. referred to these two model classes, DDPMs and SMLDs, together as {\bf score-based generative models}. Both DDPMs and SMLDs can be expanded to scenarios encompassing an infinite number of time steps or noise levels, where perturbation and denoising procedures are characterized as solutions to stochastic differential equations (SDEs). This extended framework is referred to as a score-based SDE model ({\bf ScoreSDE}) \cite{Song-ICLR-2021}. By interpreting the forward diffusion processes \eqref{discrete_DDPM} and \eqref{SMLD} as the discretization of an underlying SDE, we express the continuous-time limit as follows:
\begin{equation}
    \label{forward}
    d\mathbf{x}_t = f(\mathbf{x}_t, t) dt + g(t) d\mathbf{w}_t,
\end{equation}
where  $f(\mathbf{x}, t)$ represents the \textbf{drift term}, governing the deterministic evolution of the process, $g(t)$ denotes the \textbf{diffusion coefficient}, often defined in terms of the noise schedule, and $d\mathbf{w}_t$ indicates a \textbf{Wiener process}, modeling stochastic perturbations.
This generative framework generalizes previous score-based generative models by incorporating continuous-time SDEs and provides a mathematically flexible method to describe diffusion processes and generative modeling in continuous time.

\medskip

In \cite{Levy-2023, Rembert-2024}, the authors explored the limitations and shortcomings of the classical ScoreSDE models equipped with Brownian motion (BM). When training data have unequal representation across modes, traditional diffusion models may fail to generate samples that properly represent all modes, focusing instead on dominant modes. In addition, generated samples might lack variety, producing outputs that are too similar or fail to explore the data distribution fully. These limitations might arise due to the nature of BM with a light-tailed distribution and independent increments. Moreover, BM relies on Gaussian noise; hence, it may struggle to model data distributions with heavy tails effectively (e.g., outliers or rare, extreme variations), whereas the independence assumption limits the ability to encode dependencies or correlations across time steps, which may be necessary to model complex, multimodal distributions accurately. To overcome this limitation, researchers have investigated L\'{e}vy processes and fractional BM (fBm) as potential alternatives to BM.

\medskip

Yoon et al. \cite{Levy-2023} extended classical ScoreSDE models by replacing the underlying BM with a L\'{e}vy process, a stochastic process characterized by independent and stationary increments (i.e., the differential $d\mathbf{w}_t$ in Equation \eqref{forward} is replaced by $d L^{\alpha}_t$, where $L^{\alpha}_t$ is an $\alpha$-stable L\'{e}vy  process). Stable processes necessarily have a stability index of $\alpha$, which lies in the range $(0, 2]$. When $\alpha = 2$, the process $L^{\alpha}_t$ reduces to the classical BM and, therefore, necessarily has continuous paths. These models enable more efficient exploration of the conformational space and are applied to protein generation. Motivated by \cite{Levy-2023}, the authors of \cite{Levy-protein-2024} presented innovative L\'{e}vy–It\={o} diffusion models that integrate Markovian and non-Markovian dynamics, incorporating fractional SDEs and L\'{e}vy distributions, with applications in protein generation.

\medskip

An alternative for the driving process is fBm, a generalization of the standard BM that introduces long-range dependencies and controlled roughness. The authors of \cite{Gabriel-GFDM-2024} presented the first continuous-time score-based generative model that employs fractional diffusion processes to govern its dynamics 
(i.e., the differential $d\mathbf{w}_t$ in Eq. \eqref{forward} is replaced by $d B^{H}_t$, where $B^{H}_t$ is a fBm, which features correlated increments and is characterized by the Hurst index $H \in (0,1)$ with $H=1/2$ corresponding to classical BM). Its precise definition is given in Definition \ref{defn-fbm}. Moreover, BM and L\'{e}vy processes are semimartingales, meaning they can be decomposed into a sum of a local martingale and a finite variation process, enabling It\={o} calculus for stochastic integration \cite{PD-2019}. A challenge in dealing with fBm is the nonsemimartingale nature, invalidating the use of classical It\={o} integrals. To maintain tractable inference and learning, the authors used a recently popularized Markov approximation of fBm (MA-fBm). They derived its reverse-time model, leading to the development of generative fractional diffusion models (GFDMs). In particular, SDEs driven by fBm are well-suited for capturing systems with temporal dependencies because they account for memory effects and correlations over time. Additional studies focusing on fBm are found in \cite{PD-2019, Rembert-2024}.

\medskip

This work proposes a novel fractional diffusion-based generative model for protein design, ProT-GFDM, employing the mathematical framework of MA-fBm. Unlike traditional diffusion models that rely on BM, the proposed approach incorporates long-range dependencies and super-diffusive behavior, enabling more expressive and controllable generative dynamics. Compared to prior diffusion models \cite{Song-ICLR-2021}, the proposed method achieves superior sample diversity and fidelity. This work also explores the effect of various noise schedules, including an alternative cosine noise schedule \cite{Alexander-2021} that provides a distinct approach to noise control and influences generative behavior. Furthermore, this work systematically evaluates the adaptability of the proposed model to stochastic and deterministic solvers, demonstrating that fractional diffusion models can be effectively integrated with SDE solvers and ordinary differential equation (ODE) solvers, expanding their applicability in generative modeling. These contributions establish the approach as a flexible and robust framework for advancing generative protein design.

\medskip

The paper is organized as follows. Section \ref{Background} discusses the background of ScoreSDE, laying the theoretical foundation and presenting the relevant prior work. Next, Section \ref{Protein-representation} explores protein-structure image generation, where the protein backbone is represented by an $\alpha$-carbon distance map. Section \ref{core-GFDM} focuses on the core contribution of this work and is subdivided into parts that explore the fractional driving noise and its Markov approximation in subsection \ref{Fractional-approximation}, generative fractional diffusion models in \ref{model}, and augmented score-matching techniques in \ref{ScoreMatching-section}. This section also includes a detailed explanation of sampling methods in \ref{sampler}, presenting multiple approaches, such as SDE and ODE solvers, for efficient sampling. Following this, Section \ref{Experimental-results} presents the experimental results, highlighting the performance and validation of the proposed methods. Section \ref{Conclusion} presents the conclusions and future directions for research. Finally, Section \ref{Notational-conventions} provides the notational conventions, clarifying the symbols and terminology used throughout the paper.

%%%%%%%%%%%%%%%%%%%%%%%%%%%%%%%%%%%%%%%%%%%%%
\section{Background of the ScoreSDE Framework}
\label{Background}
To establish the connection between score-based generative models and SDEs, we examine the discrete-time formulation of DDPMs, as given in \eqref{discrete_DDPM}. We define a discrete step size $ \Delta t = \frac{1}{N} $, where $N$ represents the total number of discrete steps in the diffusion process. To facilitate the transition to a continuous formulation, we introduce an auxiliary noise schedule $\{\bar{\beta}_i\}_{i=1}^{N}$ where $\beta_i = \frac{\bar{\beta}_i}{N}$. By expressing $\beta_i$ in terms of a continuous function, we obtain the following:
$$
\beta_i = \bar{\beta}_i \cdot \frac{i}{N} = \beta\left( \frac{i}{N} \right) \cdot \frac{1}{N} = \beta(t + \Delta t)\Delta t,
$$
where this work assumes that $\bar{\beta}_i \to \beta(t)$ as $ N \to \infty $, which is a continuous time function for $ 0 \leq t \leq 1 $. We let
$$
\mathbf{x}_i = \mathbf{x}\left( \frac{i}{N} \right) = \mathbf{x}(t + \Delta t), \quad \boldsymbol{\varepsilon}_i = \boldsymbol{\varepsilon}\left( \frac{i}{N} \right) = \boldsymbol{\varepsilon}(t + \Delta t).
$$
Hence, the Taylor expansion of $\sqrt{1-x}$ for approximation yields
\begin{align*}
\mathbf{x}_i &= \sqrt{1 - \beta_i}\mathbf{x}_{i-1} + \sqrt{\beta_i}\boldsymbol{\varepsilon}_{i-1}\\
\mathbf{x}_i &= \sqrt{1 - \frac{\bar{\beta}_i}{N}}\mathbf{x}_{i-1} + \sqrt{\frac{\bar{\beta}_i}{N}}\boldsymbol{\varepsilon}_{i-1}\\
\mathbf{x}(t + \Delta t) &= \sqrt{1 - \beta(t + \Delta t)\Delta t} \cdot \mathbf{x}(t) + \sqrt{\beta(t + \Delta t)\Delta t} \cdot \boldsymbol{\varepsilon}(t)\\
\mathbf{x}(t + \Delta t) &\approx \left( 1 - \frac{1}{2}\beta(t + \Delta t)  \Delta t \right) \cdot \mathbf{x}(t) + \sqrt{\beta(t + \Delta t)\Delta t} \cdot \boldsymbol{\varepsilon}(t)\\
\mathbf{x}(t + \Delta t) &\approx \mathbf{x}(t) - \frac{1}{2}\beta(t)\Delta t \, \mathbf{x}(t) + \sqrt{\beta(t)\Delta t} \cdot  \boldsymbol{\varepsilon}(t).
\end{align*}
As $ N \to \infty $, the discrete Markov chain in Eq. \eqref{discrete_DDPM} converges to the following SDE:
\begin{equation}
\label{DDPM-sde}
d\mathbf{x}_t = -\frac{1}{2}\beta(t)\mathbf{x}_t \, dt + \sqrt{\beta(t)} \, d\mathbf{w}_t. 
\end{equation}
Therefore, the DDPM forward update iteration can be written equivalently as an SDE. The variance \( \beta(t) \) remains bounded as \( t \to \infty \) because \( \beta_i \) is constrained, ensuring a controlled noise increase that prevents the data from being completely overwhelmed. This bounded nature of variance is why it is referred to as the \textbf{variance-preserving (VP) SDE}.

\medskip

{The SMLD \eqref{SMLD} framework has no explicit forward diffusion process as in DDPMs. However, we can approximate it using $N$ discrete noise scales. Under this formulation, the recursive updates naturally follow a Markov chain:
\[
\mathbf{x}_i = \mathbf{x}_{i-1} + \sqrt{\sigma_i^2 - \sigma_{i-1}^2}\mathbf{z}_{i-1}, \quad i = 1, \cdots, N.
\]
Assuming that the limit $\{ \sigma_i \}_{i = 1}^{N}$ becomes the continuous-time function $ \sigma(t) $ for $ t \in [0, 1] $, $ \mathbf{z}_i $ becomes $ \mathbf{z}(t) $, and $\{ \bold{x}_i \}_{i = 1}^{N}$ becomes $\bold{x}_t$, where $\bold{x}_i = \bold{x}(\frac{i}{N})$, then we obtain the following:
\begin{align*}
\bold{x}(t + \Delta t) &= \bold{x}(t) + \sqrt{\sigma^2(t +\Delta t ) - \sigma^2(t)} \cdot \mathbf{z}(t) \approx \bold{x}(t) + \sqrt{\frac{d}{d t}\sigma^2(t) \Delta t} \cdot \mathbf{z}(t)
\end{align*}
At the limit when $\Delta t \to 0$, the equation converges to
\begin{equation}
\label{SMLD-sde}
d\mathbf{x} = \sqrt{\frac{d}{dt} \sigma^2(t)} \, d\mathbf{w}.
\end{equation}
Unlike the VP SDE, where noise remains bounded, the diffusion term in \eqref{SMLD-sde} grows exponentially with time. Hence, the variance of the data distribution diverges as $t \to \infty$, which is why this formulation is called {\bf variance exploding SDE}.

\medskip

Although SMLD and DDPM have demonstrated strong generative capabilities, they suffer from limitations, including slow sampling due to their iterative denoising steps and difficulties in likelihood evaluation. To address these challenges, the ScoreSDE model extends diffusion models to a continuous-time framework, offering critical advantages, including efficient likelihood computation via its connection to a probability-flow ODE (PF-ODE) and greater flexibility in sampling methods. The model is formulated as a stochastic process governed by a pair of SDEs: the forward and reverse-time SDEs.

\medskip

\noindent\textbf{Forward SDE (data-to-noise):} A general framework for many score-based generative models is where noise is injected into the data distribution $p_{\text{data}} \equiv p_0$ via a forward SDE with more general drift and diffusion coefficients:
\begin{equation}
\label{general-SDE}
d\bold{x}_t  = f(\bold{x}_t ,t) dt  + g(\bold{x}_t, t) d \bold{w}_t, \quad \bold{x}_0 \sim p_0,
\end{equation}
where $t \in [0,T]$ resides in the continuous-time domain and $\bold{w}_t$ is the standard Wiener process (a.k.a. BM). The drift coefficient $f(\cdot, t): \mathbb{R}^d \to \mathbb{R}^d$ suggests how the diffusion particle should flow to the lowest energy state, whereas the diffusion coefficient $G(\cdot, t): \mathbb{R}^d \to \mathbb{R}^{d}$ describes how the particle would randomly walk from one position to another, determining the strength of the random movement (i.e., the intensity of the random fluctuation). 

\medskip

\noindent\textbf{Reverse-time SDE (noise-to-data):} To generate new samples, the reverse-time SDE inverts this process, starting from noise and gradually reconstructing structured data by removing perturbations. A crucial result from \cite{Anderson-1982} reveals that the reverse of a diffusion process is a diffusion process, running backward in time and governed by the {\bf reverse-time SDE}:
\begin{equation}
\label{reverse-time-sde}
    d\bold{x}_t = \left\{ f(\mathbf{x}_t, t) - g(\mathbf{x}_t, t)^2  \nabla_{\mathbf{x}} \log p_t(\mathbf{x}_t) \right\} dt + g(\mathbf{x}_t, t) d\mathbf{\bar{w}}_t,
\end{equation}
where $\mathbf{\bar{w}}_t$ denotes the standard Wiener process in reverse time (i.e. time flows backward from $T$ to $0$). When the reverse-time SDE is well-constructed, we can simulate it with numerical approaches to generate new samples from $p_0$.

\medskip

When discretizing SDEs, the step size $\Delta t$ is limited by the randomness of the stochastic process \cite{Kloeden-1992}. A large step size (consequently, a small number of steps) often causes nonconvergence, especially in high-dimensional spaces, and the numerical scheme might fail to capture the correct dynamics, leading to a poor approximation of the solution’s distribution and negatively affecting the capability of high-quality sample generation. However, when the step size is small, the iterative process that generates high-quality samples, often involving hundreds or thousands of denoising steps, makes the generation process computationally expensive and slow. 

Additionally, SDE solvers do not offer a method to compute the exact log-likelihood of score-based generative models. To address those limitations, Song et al. \cite{Song-ICLR-2021} introduced a sampler based on ODEs, called the {\em PF ODE}, by converting any SDE into an ODE without changing its marginal distributions $\{ p_t(\bold{x})\}_{t \in [0, T]}$. Song et al. suggested that each SDE has a corresponding PF-ODE that produces deterministic processes, sampling from the same distribution as the SDE at each timestep. 

\medskip

\noindent\textbf{Probability-Flow ODE:} The PF-ODE corresponding to Eq.~\eqref{general-SDE} follows the general form:
\begin{equation}
\label{pf-ode}
    d\bold{x}_t = \left\{ f(\mathbf{x}, t) - \frac{1}{2} \nabla \cdot \left[ G(\mathbf{x}, t) G(\mathbf{x}, t)^\top \right] - \frac{1}{2} G(\mathbf{x}, t) G(\mathbf{x}, t)^\top \nabla_{\mathbf{x}} \log p_t(\mathbf{x}_t) \right\} dt.
\end{equation}
Appendix D.1 of \cite{Song-ICLR-2021}  provides a detailed derivation of transforming the reverse SDE into the PF-ODE. The proof is based on the Fokker--Planck (or the Kolmogorov forward) equation. Trajectories obtained by solving the PF-ODE \eqref{pf-ode} have the same marginal distributions as the SDE trajectories \eqref{reverse-time-sde}. The critical component in \eqref{pf-ode} is the score function $\nabla_{\mathbf{x}} \log p_t(\mathbf{x}_t)$, which is approximated by neural networks. This approximation creates a parallel with neural ODEs \cite{NeuralODE-2018}, inherits all properties of neural ODEs, and enables sampling via ODE solvers and the precise computation of log-likelihoods. The PF-ODE also eliminates the stochasticity from the reverse process and allows for faster sampling in generative diffusion models while maintaining high-quality output \cite{Song-ICLR-2021}. Although the reverse process becomes deterministic, multiple diverse outputs can still be generated by starting from different random noise vectors because the ODE's initial conditions vary accordingly.

\begin{remark}
The equations (SDEs or ODEs) above describe how a system changes or evolves over time. Numerical solvers are algorithms that take these equations and produce a series of points that form a “trajectory” or “path” of how the system evolves by breaking time into small steps and applying approximate formulas. This trajectory is an approximation rather than an exact representation; however, it is typically sufficient to capture and understand the behavior of a dynamic system. In generative modeling, we generate new images by numerically solving Eqs.~\eqref{reverse-time-sde} and \eqref{pf-ode}. 
\end{remark}

%\cleardoublepage
%%%%%%%%%%%%%%%%%%%%%%%%%

%%%%%%%%%%%%%%%%%%%%%%%%%
\section{Protein Backbone Representation: $\alpha$-carbon Distance Map}
\label{Protein-representation}
Proteins are macromolecules comprising linear chains of amino acids. Each amino acid residue contributes three principal backbone atoms: nitrogen ($N$), alpha carbon ($C_{\alpha}$), and carbonyl carbon ($C$), as well as carbonyl oxygen ($O$). These repeating units form the polypeptide backbone $N - C_{\alpha} - C - O$, with side chains branching from the $C_{\alpha}$ atom. Various computational methods have been developed to predict protein function. These methods can be categorized into four groups based on the information they employ (there is overlap and correlation between them). Sequence-based methods include BLAST \cite{Altschul},  DEEPred \cite{Rifaioglu-2019}, and DeepGOPlus \cite{Kulmanov-2020}, etc.). The 3D structure-based methods include DeepFRI \cite{Vladimir-2021}, LSTM-LM \cite{Hochreiter-1997}, and HEAL \cite{Zhonghui-2023}. In addition, the PPI network-based methods include GeneMANIA \cite{Mostafavi-2008} and deepNF \cite{Gligorijevic-2018}, among others. Finally, the hybrid information-based methods include the CAFA challenge \cite{Zhou-2019}. A comprehensive review and comparison of those methods is presented in \cite{Baohui-2024}. Diffusion models have recently been adopted into protein engineering applications and have demonstrated strong performance in generating novel protein structures and sequences \cite{Mardikoraem-2023, Wang-2022, Lee-2023}.

\medskip

The dataset was derived from the Protein Data Bank (PDB) \cite{Frances-1978}, a globally recognized repository archiving detailed 3D structural biomolecule data, including proteins, nucleic acids, and complex assemblies. This benchmark protein dataset was established in 1971, serving as a vital resource for researchers in structural biology, bioinformatics, and related fields and facilitating advances in drug discovery, molecular biology, and biotechnology. The PDB offers free access to the structural data submitted by scientists worldwide in a standardized format that supports computational analysis and visualization.

\medskip

Figure \ref{protein-process} illustrates the data preparation process, beginning with extracting atomic coordinates from a protein structure. The 3D model on the left represents the atomic arrangement of 101M sperm whale myoglobin, obtained from the PDB. The table in the center displays a segment of the PDB file, detailing atomic positions in a structured format. Using the extracted $(X, Y, Z)$ coordinates, the Euclidean distances between pairs of $C_{\alpha}$ atoms are computed to construct the distance matrix, visualized as a heatmap (right). The color gradient (yellow to blue) encodes distance magnitudes, with the diagonal representing zero distance, corresponding to self-comparisons. An invariant representation is necessary to analyze protein structures independent of their spatial orientation. Proteins can adopt arbitrary poses in 3D space; hence, direct coordinate-based representations are sensitive to translation and rotation. Computing pairwise distances between $C_{\alpha}$ atoms yields a structured representation, the distance matrix, which remains invariant to such transformations while preserving critical structural information. This representation is useful because it retains the sequential ordering of amino acids, defined by the $N$-terminus to $C$-terminus directionality, ensuring uniqueness, and it preserves sufficient information to recover the structure \cite{Anand-2018}.

\begin{figure}[H]
\centering
\includegraphics[width=14cm]{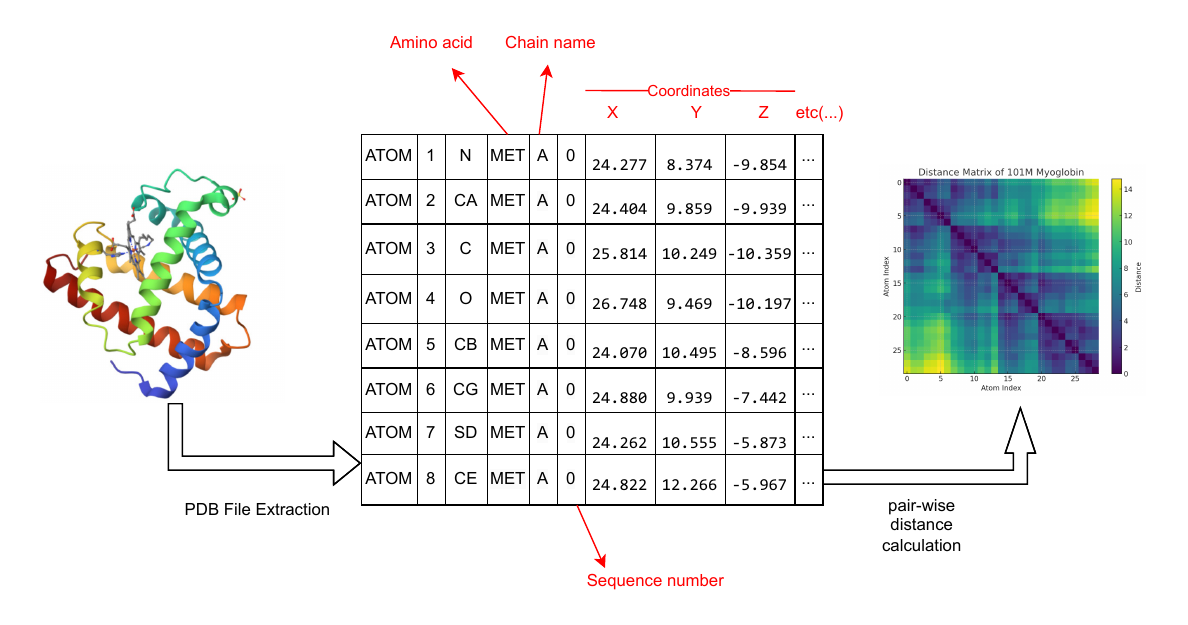}
\caption{Protein structure representation and distance matrix.}
\label{protein-process}
\end{figure}

Following the generation of the $C_{\alpha}$ distance map using a trained model, the 3D structure of the protein backbone can be reconstructed from the resulting image. Recovering or folding the 3D structure of a protein from the $\alpha$-carbon distance matrix is a classical inverse problem in structural biology, which many studies have investigated. In the existing literature on 3D protein structure recovery, notable methods include the hidden Markov model (HMM)–based approaches FB5-HMM \cite{Thomas-2006} and TorusDBN \cite{Wouter-2008}, the multiscale torsion angle GAN, 3DGAN \cite{Jiajun-2016}, and a full-atom GAN. 

Moreover, Anand et al. \cite{Anand-2018} introduced a GAN-based approach to generate sequence-sensitive, fixed-length protein structure fragments. Using these GANs, they produced $C_{\alpha}$ pairwise distance maps and used the alternating direction method of multipliers (ADMM) alongside Rosetta minimization \cite{Rohl-2004} to reconstruct the full protein structures from the derived distance constraints. The ADMM is an algorithm proposed in \cite{Stephen-2011} that solves convex optimization problems by breaking a complicated one into smaller pieces, each of which is easier to handle. The ADMM follows a decomposition-coordination strategy, in which solutions to smaller local subproblems are synchronized to solve a broader global problem. Building on \cite{Anand-2018}, Anand et al. \cite{Anand-2019} further refined the method by training deep neural networks to recover and refine full-atom pairwise distance matrices accurately for fixed-length fragments.

%\cleardoublepage
%%%%%%%%%%%%%%%%%%%%%%%%%

%%%%%%%%%%%%%%%%%%%%%%%%%
\section{Generative Fractional Diffusion Models}
\label{core-GFDM}
This section introduces the driving process, the fBm, along with its Markov approximation. Next, the generative fractional diffusion models are presented, followed by score-matching techniques designed to estimate the corresponding score function. Finally, this section discusses the SDE and ODE solvers used to model the data generation and sampling process in detail.

\subsection{Fractional driving noise and its Markov approximation}
\label{Fractional-approximation}
\begin{definition}[Fractional Brownian Motion (Types I and II)]
\label{defn-fbm}
The fBm family of continuous-time \textit{Gaussian processes} is parameterized by the \textit{Hurst index} $H \in (0,1)$, which controls the smoothness and correlation of the process. It generalizes the standard BM and exhibits self-similarity and stationary increments. Two common types of fBm exist:
\begin{itemize}
    \item \textbf{Type I:} Denoted by $W^H_t$, it contains the covariance function:
    \begin{equation*}
        \mathbb{E} \big[ W^H_t W^H_s \big] = \frac{1}{2} \big( |t|^{2H} + |s|^{2H} - |t-s|^{2H} \big).
    \end{equation*}
    
    \item \textbf{Type II:} Denoted by $B^H_t$, its covariance function is given by
    \begin{equation*}
        \mathbb{E} \big[ B^H_t B^H_s \big] = \frac{1}{(\Gamma(H + \frac{1}{2}))^2} 
        \int_{0}^{s} \big( (t - u)(s - u) \big)^{H - \frac{1}{2}} \, \mathrm{d}u, \quad s < t,
    \end{equation*}
\end{itemize}
where $\Gamma(\cdot)$ denotes the Gamma function. Type II is often referred to as the {\bf Riemann--Liouville Volterra process}, which is commonly applied in financial mathematics. The Hurst index $H$ governs the process behavior:
\begin{itemize}
    \item $H = 1/2$ corresponds to the standard BM.
    \item If $H > 1/2$, the process increments are positively correlated.
    \item If $H < 1/2$, the process increments are negatively correlated.
\end{itemize}
\end{definition}
Figure \ref{Hurst-index} illustrates how the Hurst index influences the smoothness of the fBm paths. When the Hurst index $H > 1/2$, the increments of the fBm are positively correlated, making its trajectory smoother than the standard BM. Conversely, when $H < 1/2$, the increments are negatively correlated, resulting in a rougher path compared to the BM. 
\begin{figure}[H]
\centering
\includegraphics[width=6cm]{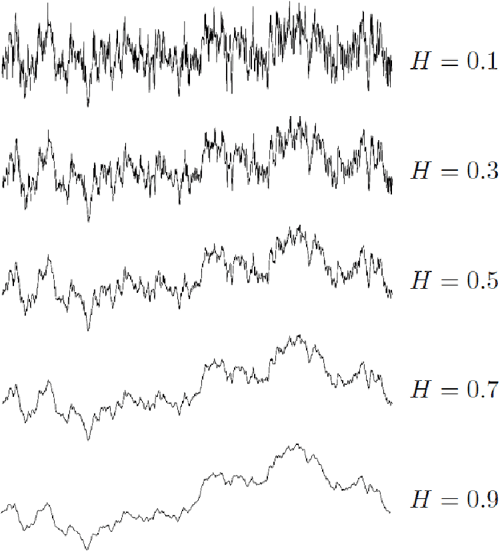}
\caption{Sample paths of fractional Brownian motion (fBm) by Hurst index $H$ values (\cite{Shevchenko-2015}).}
\label{Hurst-index}
\end{figure}

\noindent The fBm differs from standard BM in that it has long-range dependence, and its increments are not independent, making it non-Markovian. More precisely, the Markov property implies that future states depend only on the present, whereas fBm incorporates memory effects, where past values influence future behavior via the Hurst parameter. Additionally, fBm is not a semimartingale because it lacks the necessary decomposition as a local martingale and a finite variation process, restricting it from being applied directly in stochastic calculus like standard BM. This nonsemimartingale property limits its compatibility with It\={o} calculus. 

\medskip

The nature of fBm poses a challenge for score-based generative models, which typically rely on Markovian SDEs where the future state depends only on the present. In \cite{PD-2019}, the authors developed a framework for approximating fBm using Markov processes, called MA-fBm. Their critical contribution is constructing finite-dimensional, affine Markov processes approximating the behavior of fBm, which lacks the Markov property due to its long-range dependence. This approximation is crucial for the efficient simulation of fractional diffusion processes and ensures compatibility with score-based generative modeling. This approach enables the application of existing score-based generative models and well-established methods in the standard diffusion framework, facilitating the training and sampling of models driven by fractional stochastic processes.

\begin{definition}[Markov Approximation of fBm]
Choose $K \in \mathbb{N}$ Ornstein--Uhlenbeck (OU) process with speeds of mean reversion 
$\gamma_1, \ldots, \gamma_K$ and dynamics $\mathrm{d}Y^k_t = - \gamma_k Y^k_t \mathrm{d}t+ \mathrm{d}B_t$:
$$Y^k_t = Y^k_0 e^{- \gamma_k t} + \int_{0}^{t} e^{-\gamma_k (t-s)} \mathrm{d}B_s, t \ge 0,$$
where $Y^k_0 = \int_{-\infty}^{0} e^{\gamma_k s}\mathrm{d}B_s$ (Type I) and $Y^k_0 = 0$ (Type II). 
Given a Hurst index $H \in (0,1)$ and a geometric space grid $\gamma_k = r^{k-n}$ with $r > 1$ and $n = \frac{K + 1}{2}$, the process
$$\hat{B}^H_t := \sum_{k = 1}^{K} \omega_k (Y^k_t - Y^k_0), H \in (0, 1), t \ge 0,$$
is the MA-fBm with approximation coefficients $\omega_1, \ldots, \omega_K \in \mathbb{R}$.
\end{definition}

To approximate fBm $B_t^H$ with minimal errors, we choose a geometrically spaced grid  
$$
\gamma_k = (r^{1-n}, r^{2 - n}, \dots, r^{K - n}),
$$ 
where $n = \frac{K+1}{2}$ and $r > 1$. The optimal approximation coefficients $\boldsymbol{\omega} = (\omega_1, \dots, \omega_K) \in \mathbb{R}^K$, for a given Hurst index $H \in (0,1)$ and terminal time $T > 0$, are obtained by minimizing the $L^2(P)$-error:  
\[
\varepsilon(\boldsymbol{\omega}) := \int_{0}^{T} \mathbb{E} \Big[ \big( B_t^H -  \hat{B}^H_t\big)^2 \Big] \mathrm{d}t.
\]  
The optimal coefficients satisfy the following closed-form system:  
\[
\bold{A} \boldsymbol{\omega} = \bold{b},
\]  
where matrix $\bold{A}$ and vector $\bold{b}$ are given for the \textbf{Type I fBm approximation} by  
$$
\bold{A}_{i,j} := \frac{2T + \frac{e^{-\gamma_iT} - 1}{\gamma_i} + \frac{e^{-\gamma_jT}-1}{\gamma_j}}{\gamma_i + \gamma_j},
$$
$$
\bold{b}_{k} := \frac{2T}{\gamma^{H + \frac{1}{2}}_k}
- \frac{T^{H + \frac{1}{2}}}{\gamma_k \Gamma(H + \frac{3}{2})}
+ \frac{
e^{-\gamma_k T} - Q(H + \frac{1}{2}, \gamma_k T) e^{\gamma_k T}
}{\gamma^{H + \frac{3}{2}}_k},
$$
where $Q(z, x)$ represents the \textbf{regularized upper incomplete gamma function}:  
$$
Q(z, x) = \frac{1}{\Gamma(z)} \int_{x}^{\infty} t^{z - 1} e^{-t} \mathrm{d}t.
$$
For \textbf{Type II fBm approximation}, the corresponding expressions for $\bold{A}$ and $\bold{b}$ are  
$$
\bold{A}_{i,j} := \frac{T + \frac{e^{-(\gamma_i + \gamma_j)T} - 1}{\gamma_i + \gamma_j}}{\gamma_i + \gamma_j},
$$
$$
\bold{b}_{k} := \frac{T}{\gamma^{H + \frac{1}{2}}_k} P\big( H + \frac{1}{2}, \gamma_kT \big) - 
\frac{H + \frac{1}{2}}{\gamma^{H + \frac{3}{2}}_k} P \big( H + \frac{3}{2}, \gamma_kT \big),
$$
where $P(z, x)$ denotes the \textbf{regularized lower incomplete gamma function}:  
$$
P(z, x) = \frac{1}{\Gamma(z)} \int_{0}^{x} t^{z - 1} e^{-t} \mathrm{d}t.
$$
\noindent For further discussion on practical considerations in selecting geometric sequences $\gamma_k$ and the time horizon for optimizing the coefficients $\boldsymbol{\omega}$, see \cite{Rembert-2024}. The experiments in \textbf{Section \ref{Experimental-results}} focus on the Type II fBm approximation.

%\cleardoublepage

\subsection{Fractional noise-driven generative diffusion model}
\label{model}
The framework is based on a simplified SDE where the diffusion coefficient does not depend on $\bold{X}_t$, that is, the diffusion scale $g$ in \eqref{general-SDE} is only time-dependent, rather than time- and state-dependent:
\begin{equation}
\label{GFDM}
\mathrm{d}\bold{X}_t  = u(t)\bold{X}_t \mathrm{d}t  + g(t) \mathrm{d} \boldsymbol{\hat{B}}^H_t, \quad \bold{X}_0 \sim p_0.
\end{equation}
The OU processes $Y^k_t$, $k = 1, \ldots, K$ approximate $\boldsymbol{\hat{B}}^H_t$ satisfying the dynamics $\mathrm{d}Y^k_t = - \gamma_k Y^k_t \mathrm{d}t+ \mathrm{d}B_t$; thus, we have the following:
\begin{equation}
\label{MA-FBM}
\mathrm{d} \boldsymbol{\hat{B}}^H_t = - \sum_{k = 1}^{K} \omega_k \gamma_k Y^k_t \mathrm{d}t 
+ \sum_{k = 1}^{K} \omega_k \mathrm{d} \bold{B}_t.
\end{equation}
Rewriting the dynamics \eqref{GFDM} with \eqref{MA-FBM} yields
\begin{equation}
\mathrm{d}\bold{X}_t  = \Big[ u(t)\bold{X}_t - g(t) \sum_{k = 1}^{K} \omega_k \gamma_k Y^k_t   \Big] \mathrm{d}t  + \sum_{k = 1}^{K} \omega_k g(t) \mathrm{d} \bold{B}_t.
\end{equation}
Considering the forward and OU processes defining the driving noise $\hat{B}^H_t$, we have an augmented vector of the correlated process $Z \equiv (X, Y^1, \ldots, Y^K) = (Z_t)_{t \in [0,T]}$, driven by the same BM
\begin{equation}
\label{classical-GFDM-forward}
\mathrm{d} \bold{Z}_t = \bold{F}(t) \bold{Z}_t \mathrm{d}t + \bold{G}(t) \mathrm{d}\bold{B}_t, \quad t \in [0, T],
\end{equation}
where
%\begin{align*}
%\bold{F}(t) = 
%\begin{pmatrix}
%u(t) \bold{X}_t & -g(t) \omega_1 \gamma_1 Y^1 & -%g(t) \omega_2 \gamma_2 Y^2 & \ldots & -g(t) \omega_K %\gamma_K Y^K\\
%0 & - \gamma_1 Y^1 & 0 & \ldots & 0\\
%0 & 0 & - \gamma_2 Y^2 & \ldots & 0\\
%\vdots & \vdots & \vdots & \ddots & \vdots \\
%0 & 0 & 0 & \ldots & - \gamma_K Y^K
%\end{pmatrix}_{(K+1) \times  (K+1)}
%\end{align*}
\begin{align*}
\bold{F}(t) = 
\begin{pmatrix}
u(t) & -g(t) \omega_1 \gamma_1  & -g(t) \omega_2 \gamma_2  & \ldots & -g(t) \omega_K \gamma_K \\
0 & - \gamma_1  & 0 & \ldots & 0\\
0 & 0 & - \gamma_2  & \ldots & 0\\
\vdots & \vdots & \vdots & \ddots & \vdots \\
0 & 0 & 0 & \ldots & - \gamma_K \\
\end{pmatrix}_{(K+1) \times  (K+1)}
\end{align*}
and
\begin{align*}
\bold{G}(t) = 
\begin{pmatrix}
\sum_{k = 1}^{K} \omega_k g(t) & 0 & 0 & \ldots & 0  \\
0 & 1 & 0  & \ldots & 0  \\
0 & 0 & 1 & \ldots & 0  \\
\vdots & \vdots & \vdots & \ddots & \vdots \\
0 & 0 & 0 &  \ldots & 1
\end{pmatrix}_{(K+1) \times  (K+1).}
\end{align*}

\medskip

The reverse time model of the GFDM \eqref{classical-GFDM-forward} is given by the backward dynamics
\begin{align*}
\mathrm{d} \bold{Z}_t 
= \bigg\{ \bold{F}(t) \bold{Z}_t - \bold{G}(t) \bold{G}(t)^T 
\nabla_{\bold{z}} \text{log} \ p_t(\bold{Z}_t ) \bigg\}\mathrm{d}t 
+ \bold{G}(t) \mathrm{d}\bold{B}_t, \quad t \in [0, T],
\stepcounter{equation}\tag{\theequation}\label{reverse-GFDM}
\end{align*}
and its PF-ODE is given by
\begin{align*}
\mathrm{d} \bold{z}_t 
= \bigg\{ \bold{F}(t) \bold{z}_t - \frac{1}{2}\bold{G}(t) \bold{G}(t)^T 
\nabla_{\bold{z}} \text{log} \ p_t(\bold{z}_t ) \bigg\}\mathrm{d}t, \quad t \in [0, T].
\stepcounter{equation}\tag{\theequation}\label{reverse-GFDM-ode}
\end{align*}

\begin{figure}[H]
\centering
\includegraphics[width=14cm]{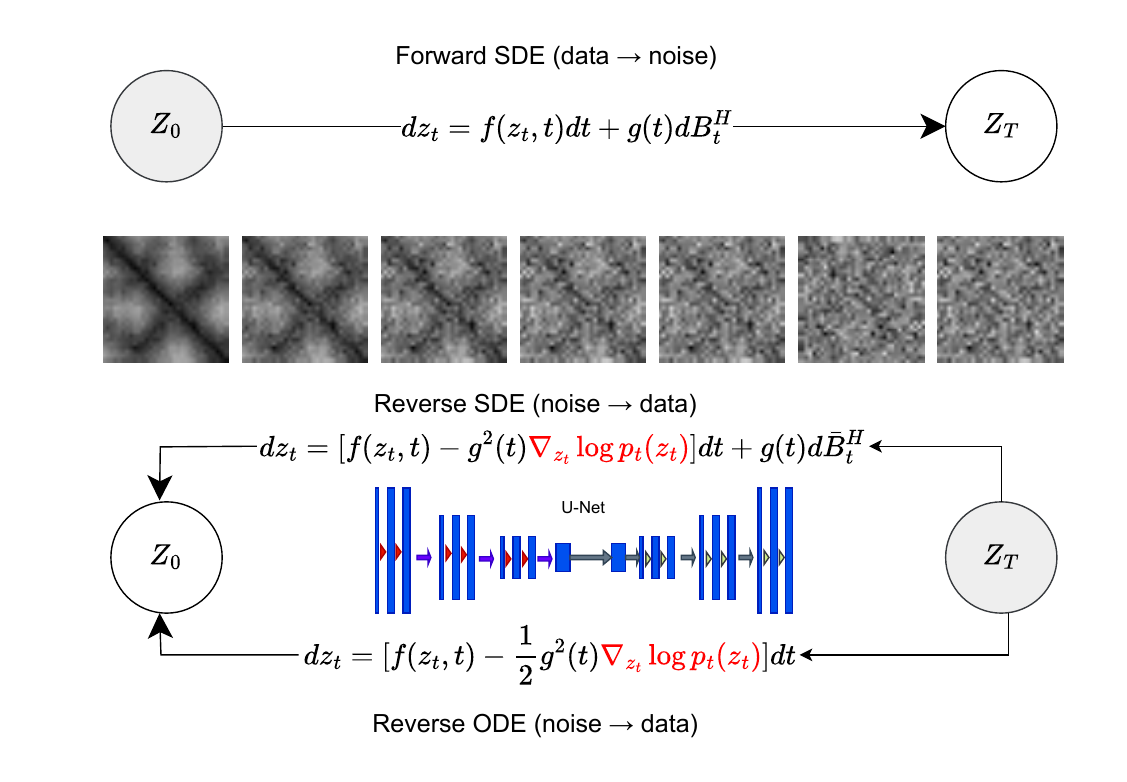}
\caption{Overview of score-based generative modeling through stochastic and ordinary differential equations (SDEs and ODEs).}
\label{fig-model-20250309}
\end{figure}

Figure \ref{fig-model-20250309} breaks down the process of the ScoreSDE model, illustrating its critical components and their interactions. The ScoreSDE employs SDEs to model data generation as a diffusion process. The forward SDE progressively perturbs actual data into a Gaussian noise distribution. In contrast, the reverse SDE inverts this process by iteratively denoising the samples, guided by a learned score function $\nabla_{z_t} \log p_t(z_t)$. Alternatively, a reverse ODE can be applied for a deterministic sampling approach, improving stability and efficiency. A U-Net architecture is commonly employed to parameterize the score function $\nabla_{z_t} \log p_t(z_t)$, facilitating precise noise estimation and denoising.

%\cleardoublepage
%%%%%%%%%%%%%%%%%%%%%%%%%

%%%%%%%%%%%%%%%%%%%%%%%%%
\subsection{Augmented score-matching technique}
\label{ScoreMatching-section}
The score function (the gradient of the log-probability density $p_{\text{data}}$), $\nabla_{\bold{x}} \log p_t(x_t)$, which appears in \eqref{reverse-time-sde} or \eqref{pf-ode}, guides the transformation of pure noise into coherent images by supplying directional information on how to increase the data likelihood at each step of the reverse diffusion process. {\bf Score matching} is a nonlikelihood-based method to perform sampling on an unknown data distribution and aims to address many of the limitations of likelihood-based methods (e.g., VAE and autoregressive models) and adversarial methods (GANs). A good estimation of the score function is crucial because it directly determines the quality, realism, and fidelity of the data generated using score-based generative models. In contrast, an inaccurate score estimation can lead to poor generation results, including artifacts, mode collapse, or failure to capture an accurate data distribution.
%The score function guides the reverse diffusion process (or reverse stochastic differential equation) by indicating how to transform noise into structured, high-probability data samples. 

\medskip

In practice, the score function is learned using a neural network $\mathbf{s}_{\theta}(\mathbf{x})$ parameterized by $\theta$, obtained by minimizing the Fisher divergence between the true score function and the network:
\begin{equation}
\label{SM-original}
J(\theta) = \mathbb{E}_{p_{\text{data}}(\mathbf{x})} \left[ \left\| \mathbf{s}_{\theta}(\mathbf{x}) - \nabla_{\mathbf{x}} \log p_{\text{data}}(\mathbf{x}) \right\|_2^2 \right].
\end{equation}
The problem is that $p_{\text{data}}$ is never accessible. Various methods can estimate the score function without knowledge of the ground-truth data score. 

\medskip

\noindent\textbf{Implicit score matching (ISM)} Hyv\"{a}rinen and Dayan \cite{Aapo-Peter-2005} introduced the implicit score matching method and demonstrated its equivalence to $J(\theta)$ in \eqref{SM-original}, up to a constant, under mild regulatory conditions:
\begin{equation}
\label{ISM}
J_{ISM}(\theta) = \mathbb{E}_{p_{\text{data}}(\mathbf{x})} \Big[ \frac{1}{2} \lVert \mathbf{s}_{\theta}(\mathbf{x}) \rVert^2_2 + \text{tr}(\nabla_{\mathbf{x}} \mathbf{s}_{\theta}(\mathbf{x})) \Big] + \text{constant},
\end{equation}
where $\text{tr}(\nabla_{\mathbf{x}} \mathbf{s}_{\theta}(\mathbf{x}))$ denotes the trace of the Hessian matrix of the log-probability with respect to $\bold{x}$. Equation \eqref{ISM} is derived by simplifying \eqref{SM-original} using the generalized integration by parts (mutidimensional) technique. Minimizing $J_{ISM}(\theta)$ does not require the true target scores $\nabla_{\mathbf{x}} \log p_{\text{data}}(\mathbf{x})$, and we only need to compute an expectation with respect to the data distribution, which can be implemented using finite samples from the dataset using the Markov chain Monte Carlo method.

\medskip

\noindent\textbf{Denoising score matching (DSM)} An alternative approach is DSM, proposed by Vincent in \cite{Vincent-2011}. The DSM approach introduces noise perturbations to the data $\mathbf{x}$, creating a smoothed version of the data distribution $\tilde{\mathbf{x}}$. Instead of directly modeling the original distribution, the model learns the score function of the corrupted distribution. Given noisy data $\tilde{\mathbf{x}} = \mathbf{x} + \sigma \boldsymbol{\epsilon}$, where $\sigma$ controls noise intensity, the model minimizes the following DSM objective:
\begin{align*}
J_{DSM}(\theta) &= 
\mathbb{E}_{q_{\sigma}(\tilde{\mathbf{x}}, \mathbf{x})}
\left[ \frac{1}{2} \left\| \mathbf{s}_{\theta}(\tilde{\mathbf{x}}) - \nabla_{\tilde{\mathbf{x}}} \log q_{\sigma}(\tilde{\mathbf{x}}|\mathbf{x}) \right\|_2^2 \right]\\
&= \mathbb{E}_{q_{\sigma}(\tilde{\mathbf{x}}, \mathbf{x})}
\left[ \frac{1}{2} \left\| \mathbf{s}_{\theta}(\tilde{\mathbf{x}}) 
- \frac{\mathbf{x} - \tilde{\mathbf{x}}}{\sigma^2} \right\|_2^2 
\right]
\end{align*}
over the joint density on original-corrupt data pairs $(\tilde{\mathbf{x}}, \mathbf{x})$, which is $q_{\sigma}(\tilde{\mathbf{x}}, \mathbf{x}) = q_{\sigma}(\tilde{\mathbf{x}}|\mathbf{x}) p_{\text{data}}(\mathbf{x})$. Moreover, $\nabla_{\tilde{\mathbf{x}}} \log q_{\sigma}(\tilde{\mathbf{x}}|\mathbf{x}) $ is not the data score, but it provides the information of the direction of moving from $\tilde{\mathbf{x}}$ back to the original $\mathbf{x}$ (that is the reason this method is called ``denoising"). 
%$$
%J_{DSM}(\theta)  = \mathbb{E}_{p_{\text{data}}(\mathbf{x})} \mathbb{E}_{p(\boldsymbol{\epsilon})} \left[ \frac{1}{2} \left\lVert \mathbf{s}_{\theta}(\mathbf{x} + \sigma \boldsymbol{\epsilon}) + \frac{\boldsymbol{\epsilon}}{\sigma^2} \right\rVert^2 \right]
%$$

\medskip

\noindent\textbf{Sliced score matching (SSM)} The loss function in \eqref{ISM} requires computing the trace, which remains intractable in high-dimensional spaces, as analyzed in \cite{James-2012}. Song et al. proposed SSM \cite{YS-B1-2019-sliced, YS-B1-2019} to scale up the computation of score matching considerably. They projected the scores in random directions, such that the vector fields of the scores of the data and model distribution become scalar fields, motivated by the fact that a one-dimensional data distribution is much easier to estimate for score matching. Then, they analyzed the scalar fields to assess the disparity between the model and data distributions. The trackable training objective can be considered a random projected version of the Fisher divergence of \eqref{SM-original}:
$$
J_{SSM}(\theta) =\mathbb{E}_{p_{\mathbf{v}}} \mathbb{E}_{p_{\text{data}}(\mathbf{x})} \left[ \frac{1}{2} \lVert \mathbf{s}_{\theta}(\mathbf{x}) \rVert^2 + \mathbf{v}^{T} \nabla_{\mathbf{x}} \mathbf{s}_{\theta}(\mathbf{x}) \mathbf{v} \right] + \text{constant},
$$
where $\mathbf{v} \sim N(0, \bold{I})$ represents the random projection direction, and $p_{\mathbf{v}}$ denotes its distribution. The SSM approach is considered one alternative to DSM that is consistent and computationally efficient.

\medskip

\noindent\textbf{Noise-conditioned score network} In the previous DSM approach, the strength $\sigma$ determines how well the corrupted data distribution $q_{\sigma}(\tilde{\mathbf{x}})$ aligns with the original distribution $p_{\text{data}}(\mathbf{x})$. A trade-off occurs in selecting $\sigma$. A larger $\sigma$ value helps capture low-density regions of the data distribution, improving the score estimation. However, if $\sigma$ is too large, the data distribution becomes excessively corrupted, making learning challenging. Conversely, a smaller $\sigma$ preserves the original data distribution more accurately but fails to perturb the data sufficiently, poorly covering low-density regions. To address this problem, Song and Ermon introduced a multiscale noise perturbation approach \cite{YS-B1-2019, YS-SE-2020}, applying noise at multiple levels simultaneously. Specifically, they defined $L$ noise-perturbed data distributions, each associated with a different noise scale $\sigma$, ordered such that  
\[
\sigma_1 > \sigma_2 > \dots > \sigma_L,
\]
with higher indices corresponding to lower noise levels. The core idea behind the noise-conditioned score network method is perturbing the data distribution with the isotropic Gaussian noise $N(0, \sigma^2_i\bold{I}), i = 1, \ldots, L$ to obtain a noise-perturbed distribution:
$$
p_{\sigma_i}(\mathbf{x}) = \int p(\mathbf{y}) N(\bold{x}; \bold{y}, \sigma^2_i\bold{I}) d \bold{y}.
$$
The next step is to train the noise-conditioned score-based model $\mathbf{s}_{\theta}(\mathbf{x}, i)$ to estimate the score function of each noise-perturbed distribution, for $i = 1, \ldots, L$:
$$
\mathbf{s}_{\theta}(\mathbf{x}, i) \approx \nabla_{\mathbf{x}} \log p_{\sigma_i}(\mathbf{x}).
$$
The training objective is the weighted sum of Fisher divergences for all noise scales:
$$
\sum_{i=1}^L \lambda(i) \mathbb{E}_{p_{\sigma_i}(\mathbf{x})} \left[ \lVert  \mathbf{s}_{\theta}(\mathbf{x}, i)  - \nabla_{\mathbf{x}} \log p_{\sigma_i}(\mathbf{x})
\rVert_2^2 \right],
$$
where $\lambda(i) \in \mathbb{R}$ denotes a positive weighting function, often set to $\lambda(i) = \sigma_i^2$.

%We can always simulate the SDE to sample from \( p_{0t}(\mathbf{x}(t) | \mathbf{x}(0)) \), and solve equation to train the time-dependent score-based model \( \mathbf{s}_{\theta}(\mathbf{x}, t) \).

\medskip

\begin{remark}
In diffusion generative models, two primary approaches are commonly employed to define the model task: {\em data input prediction models} and {\em noise prediction models}. For {\em data input prediction models}, the model is trained to directly predict the original data (e.g., the clean image) $\bold{x}_0$ from the noisy data $\bold{x}_t$. At each step in the reverse process, the model estimates the clean data conditioned on the noisy input. For {\em noise prediction models}, the model is trained to predict the added noise $\boldsymbol{\varepsilon}$ in the noisy data $\bold{x}_t = \alpha_t \bold{x}_0 + \sigma_t \boldsymbol{\varepsilon}$. This approach focuses on estimating the noise rather than the clean data.

\end{remark}

\medskip

This work applies augmented score-matching techniques to learn the score function $\nabla_{\bold{z}} \log p_t$, which appears in \eqref{reverse-GFDM}. The authors in \cite{Gabriel-GFDM-2024} suggested that one can condition the score function $\nabla_{\bold{Z}} \log p_t(\bold{Z}_t)$ on a data sample $\bold{X}_0 \sim p_0$ and on the states of the stacked vector $\bold{Y}_t^{[K]} := (\bold{Y}_1^{1}, \ldots, \bold{Y}_t^{K})$ of augmenting processes. A time-dependent score model $s_{\boldsymbol{\theta}}$ can be trained by minimizing the following augmented score-matching loss:
\begin{align*}
\mathcal{L}(\boldsymbol{\theta}) :=
\mathbb{E}_{t \sim U[0,T]} 
\Bigg\{ 
&\mathbb{E}_{\bold{X}_0, \bold{Y}_t^{[K]} }
\mathbb{E}_{\bold{X}_t | \bold{Y}_t^{[K]}, \bold{X}_0}\\
&\Big\|
s_{\boldsymbol{\theta}}\Big(\bold{X}_t - \sum_k \eta_t^k Y^k_t, t \Big)
- \nabla_{\bold{x}} \log p_{0t} (\bold{X}_t | \bold{Y}_t^{[K]}, \bold{X}_0) 
\Big\|_2^2
\Bigg\}.
\end{align*}
The weights $\eta_t^k, \ldots, \eta_t^k$ arise from conditioning $\bold{Z}_t$ on $\bold{Y}_t^{[K]}$. We assume that $s_{\boldsymbol{\theta}}$ is optimal with respect to the augmented score matching loss $\mathcal{L}$. The score model
%\begin{align*}
%S_{\boldsymbol{\theta}}(\bold{Z}_t , t) &:= 
%\bigg( s_{\boldsymbol{\theta}}\Big(\bold{X}_t - \sum_k \eta_t^k Y^k_t, t \Big), 
%-\eta_t^1s_{\boldsymbol{\theta}}\Big(\bold{X}_t - \sum_k \eta_t^k Y^k_t, t \Big),\\
%& \qquad \qquad \qquad \ldots,  -\eta_t^K s_{\boldsymbol{\theta}}\Big(\bold{X}_t - \sum_k \eta_t^k Y^k_t, t \Big) \bigg)
%\end{align*}
\begin{align*}
S_{\boldsymbol{\theta}}(\bold{Z}_t , t) := 
\bigg( s_{\boldsymbol{\theta}}\Big(\bold{X}_t - \sum_k \eta_t^k Y^k_t, t \Big), 
& -\eta_t^1s_{\boldsymbol{\theta}}\Big(\bold{X}_t - \sum_k \eta_t^k Y^k_t, t \Big), \ldots,  \\
& -\eta_t^K s_{\boldsymbol{\theta}}\Big(\bold{X}_t - \sum_k \eta_t^k Y^k_t, t \Big) \bigg)
\end{align*}
yields the optimal $L^2$ approximation of $\nabla_{\bold{z}} \text{log} \ p_t(\bold{Z}_t )$ via
\begin{equation}
\nabla_{\bold{z}} \text{log} \ p_t( \bold{Z}_t ) \approx S_{\boldsymbol{\theta}}(\bold{Z}_t , t) + 
\nabla_{\bold{Y}} \text{log} \ q_t( \bold{Y}^{[K]}_t).
\end{equation}
The random vector $\bold{Y}_t^{[K]}$ is a centered Gaussian process with a covariance matrix $\boldsymbol{\Sigma}^{\bold{y}}_t \in \mathbb{R}^{K, K}$, where $[\boldsymbol{\Sigma}^{\bold{y}}_t]_{k,l} = \mathbb{E}[Y_t^k Y_t^l]$.
We can directly calculate the related score function using 
$$
\nabla_{\bold{Y}} \text{log} \ q_t( \bold{Y}^{[K]}_t)
= - \boldsymbol{\Sigma}^{\bold{y}}_t \bold{Y}^{[K]}_t.
$$

\medskip

After training the score model $S_{\boldsymbol{\theta}}(\bold{Z}_t , t)$ using augmented score matching, we can simulate the reverse-time model by running it backward and generate samples by discretizing the following reverse-time SDE ({\bf augmented reverse-time SDE}) from $T$ to $0$: 
\begin{equation}
\mathrm{d}\bold{Z}_t = 
\bigg\{ 
\bold{F}(t) \bold{Z}_t - \bold{G}(t) \bold{G}(t)^T
\Big[ 
S_{\boldsymbol{\theta}}(\bold{Z}_t , t) + \nabla_{\bold{Z}_t} \text{log} q_t( \bold{Y}^{[K]}_t)
\Big]
\bigg\}\mathrm{d}t
+ \bold{G}(t) \mathrm{d} \bar{\bold{B}}_t, 
\stepcounter{equation}\tag{\theequation}\label{augmented-reverse-sde}
\end{equation}
or the corresponding {\bf augmented PF-ODE}:
\begin{equation}
\mathrm{d}\bold{z}_t = 
\bigg\{ 
\bold{F}(t) \bold{z}_t - \frac{1}{2} \bold{G}(t) \bold{G}(t)^T
\Big[ 
S_{\boldsymbol{\theta}}(\bold{z}_t , t) + \nabla_{\bold{z}_t} \text{log} q_t( \bold{y}^{[K]}_t)
\Big]
\bigg\}\mathrm{d}t.
\stepcounter{equation}\tag{\theequation}\label{augmented-pf-ode}
\end{equation}
Commonly, \eqref{augmented-reverse-sde} and \eqref{augmented-pf-ode} are called the {\bf diffusion SDE} and {\bf diffusion ODE}, respectively. Because diffusion ODEs have no random term, one can obtain an exact solution $\bold{x}_0$, given $\bold{x}_T$, by solving the diffusion ODEs using the corresponding numerical solvers.

%\cleardoublepage
%%%%%%%%%%%%%%%%%%%%%%%%%

%%%%%%%%%%%%%%%%%%%%%%%%%
\subsection{Sampling methods}
\label{sampler}

%\begin{algorithm}
%    \caption{Predictor-Corrector (PC) sampling}\label{your_label}
%    \begin{algorithmic}
%        \STATE  $\mathrm{train\_ANN} (f_i,w_i,o_j)$
%        \FOR{epochs = $1$ to $N$}
%            \WHILE{$(j\le m)$}
%                \STATE Randomly initialize $w_i=\{w_1,w_2,\dots,w_n\}$
%                \STATE input $o_j=\{o_1,o_2,\dots,o_m\}$ in the input layer
%                \STATE forward propagate $(f_i\cdot w_i)$ through layers until getting the predicted result $y$
%                \STATE compute $e=y-y^2$
%                \STATE back propagate $e$ from right to left through layers
%                \STATE update $w_i$
%            \ENDWHILE
%        \ENDFOR
%    \end{algorithmic}
%\end{algorithm}

In generative modeling,``sampling" refers to the function of generating new data points from the learned model, creating output that resembles the training data. Efficient sampling is critical for practically deploying these models in real-world applications, such as protein, image, and audio generation, where computational resources and time are oftentimes limited. For instance, although the DDPM generates high-fidelity samples, their practical utility is limited by slow sampling speeds. 

Researchers have significantly advanced the study of sampling theory with diffusion models. For example, in \cite{DDIM-2020}, the sampling approach was extended to the denoising diffusion implicit model (DDIM) scheme. Unlike DDPMs, which rely on a stochastic reverse diffusion process, DDIMs use a deterministic sampling process and reduce the number of sampling steps from hundreds (in DDPMs) to as few as 10 to 50 steps without sacrificing image quality. 

In the ScoreSDE model, the generation process is governed by simulating the trajectories of differential equations. Samplers for diffusion models typically discretize either the reverse-time SDE (Eq.~\eqref{reverse-time-sde}) or the PF-ODE (Eq. \eqref{pf-ode}). Numerical solvers for SDEs or ODEs are employed to transform random noise iteratively into realistic samples. The importance of sampling methods in diffusion models cannot be overstated, and these methods often trade between speed (sample efficiency) and accuracy (sample quality). 

\medskip

% The authors in \cite{Gabriel-GFDM-2024} noted that the source of qualitative differences between sampling from the ODE and the SDE remains unclear. This motivates us to conduct a comparative analysis of the performance of sampling methods based on ODEs and SDEs.

%\subsubsection{{\bf SDE solver}}
%\label{SDE solve-PC-method}
{\bf SDE solver} When numerically simulating an SDE, one typically employs various methods, such as the Euler–Maruyama scheme, stochastic Runge--Kutta (RK) method \cite{Peter-2013} or more sophisticated schemes, such as Milstein’s method \cite{Milshtein-1975}. Numerical solvers display varying behaviors in terms of approximation errors. Directly applying a standard numerical solver to an SDE may introduce varying degrees of error.

\medskip

Song et al. \cite{Song-ICLR-2021} introduced a hybrid sampling algorithm, refining the Euler–Maruyama method using a \textbf{predictor-corrector (PC) sampler}, as outlined in Algorithm \ref{PC-general}. This approach is inspired by PC methods, a class of numerical techniques commonly used for solving systems of equations \cite{Eugene-2012}. The framework was designed to reduce discretization errors in reverse-time SDEs by introducing corrective steps during sample generation. The method operates in two stages per iteration:
\begin{enumerate}
    \item \textbf{Prediction} – A numerical solver (e.g., Euler–Maruyama) estimates the sample at the next time step, serving as a ``predictor.''  
    \item \textbf{Correction} – A score-based Markov chain Monte Carlo method adjusts the estimated sample’s marginal distribution, acting as a ``corrector.''  
\end{enumerate}
\noindent By iteratively refining the sample at each time step, the PC sampler improves stability and accuracy in the generative process.

%These methods involve discretizing the time interval $[0,T]$ into small steps $\Delta t$. 
\begin{algorithm}[H]
\caption{Predictor-corrector (PC) sampling}
\label{PC-general}
\begin{algorithmic}[1]
\Require $N$: Number of discretization steps for the reverse-time SDE
\Require $M$: Number of corrector steps
\State Initialize $\mathbf{x}_N \sim p_T(\mathbf{x})$
\For{$i = N - 1$ to $0$}
    \State $\mathbf{x}_i \leftarrow$ Predictor($\mathbf{x}_{i+1}$)
    \For{$j = 1$ to $M$}
        \State $\mathbf{x}_i \leftarrow$ Corrector($\mathbf{x}_i$)
    \EndFor
\EndFor
\State \textbf{return} $\mathbf{x}_0$
\end{algorithmic}
\end{algorithm}
In general, the predictor can be any numerical solver for the reverse-time SDE with a fixed discretization strategy (e.g., a {\em reverse diffusion sampler} (Eq.~(46) in \cite{Song-ICLR-2021}) or an {\em ancestral sampling} (Eq.~(4) in \cite{Song-ICLR-2021})). The corrector is typically Markov chain Monte Carlo based, such as Langevin dynamics or Hamiltonian Monte Carlo, and solely relies on a score function. For example, Algorithm~\ref{PC-example} illustrates that the reverse diffusion SDE solver can be set as the predictor and annealed Langevin dynamics as the corrector, where $\theta^{\ast}$ denotes the optimal parameter of the networks $s_{\theta}$ and $\{ \epsilon_i \}_{i = 0}^{ N - 1}$ denotes the step sizes for Langevin dynamics.
\begin{algorithm}[H]
\caption{PC sampling (VP SDE)}
\label{PC-example}
\begin{algorithmic}[1]
\Require $N$: Number of discretization steps for the reverse-time SDE
\Require $M$: Number of corrector steps
\State $\mathbf{x}_N \sim \mathcal{N}(0, \mathbf{I})$
\For{$i = N - 1$ to $0$}
    \State $\mathbf{x}_i' \leftarrow (2 - \sqrt{1 - \beta_{i+1}})\mathbf{x}_{i+1} + \beta_{i+1}\mathbf{s}_{\theta^{\ast}}(\mathbf{x}_{i+1}, i+1)$
    \State $\mathbf{z} \sim \mathcal{N}(0, \mathbf{I})$
    \State $\mathbf{x}_i \leftarrow \mathbf{x}_i' + \sqrt{\beta_{i+1}} \mathbf{z}$ \hfill \colorbox{blue!10}{Predictor}
    \For{$j = 1$ to $M$}
        \State $\mathbf{z} \sim \mathcal{N}(0, \mathbf{I})$ \hfill \colorbox{orange!10}{Corrector}
        \State $\mathbf{x}_i \leftarrow \mathbf{x}_i + \epsilon_i \mathbf{s}_{\theta^{\ast}}(\mathbf{x}_i, i) + \sqrt{2\epsilon_i}\mathbf{z}$
    \EndFor
\EndFor
\State \textbf{return} $\mathbf{x}_0$
\end{algorithmic}
\end{algorithm}

\medskip

%%%%%%%%%%%%%%%%

% \subsubsection{{\bf ODE solver}}
% \label{ODE-solver-subsection}
{\bf ODE solver} The sampling of the continuous-time diffusion model can also be created by solving the corresponding PF-ODEs of the generative model because such an ODE has the remarkable property that its solution at each time $t$ shares the same marginal distribution as the original SDE solution at that time. Because the PF-ODE is deterministic, it often enables more sophisticated and adaptive ODE solvers. These solvers can take larger steps or apply adaptive step sizing without worrying about capturing random increments at each step. This approach might result in faster numerical methods with better convergence properties, especially when scaling up to high-dimensional problems. One potential advantage of working with the PF-ODE as an alternative is its ability to enable faster sampling. Moreover, Chen et al. \cite{Chen-pfode-2023} demonstrated the polynomial-time convergence guarantees for PF-ODEs in the context of score-based generative models, emphasizing their theoretical robustness and efficiency.

\medskip

\section{Experimental Setup and Results}
\label{Experimental-results}

\textbf{Explicit fractional forward dynamics} The framework presented in \cite{Gabriel-GFDM-2024} is not limited to a specific choice of drift and diffusion coefficients. The experiments focus on \textbf{fractional VP (FVP)} dynamics with different noise schedule values, which are given by
$$
dX_t = -\frac{1}{2} \beta(t) X_t \, dt + \sqrt{\beta(t)} \, d \tilde{B}_t^H, \quad t \in [0, T].
$$
\noindent\textbf{Noise schedule} Noise schedules (i.e., $f$ and $G$ in Eq.~\eqref{general-SDE}) in diffusion models determine the addition and removal of noise during the diffusion process, significantly influencing sampling quality, training stability, and convergence. The correct selection of the noise schedule and steps is critical to optimizing model performance. In \cite{Jonathan-2020}, the authors introduced linear or quadratic schedules because they are simple and intuitive to implement. However, some authors \cite{Alexander-2021, Kingma-VDM-2021} have criticized this approach, emphasizing that the steep decline in the initial time steps creates challenges for the neural network model during generation. Alternative noise scheduling functions with a more gradual decay have been proposed to address this problem. In particular, a cosine noise schedule \cite{Alexander-2021} improves log-likelihoods compared with the conventional linear noise schedule in diffusion models. Unlike the linear schedule, the cosine schedule ensures a smoother progression of noise levels, introducing noise more gradually at the beginning and end while accelerating the transition in the middle stages. This adaptive noise scaling preserves structural information early on while maintaining efficient denoising in later steps to improve sample quality. Table \ref{tab:NS} summarizes these two types of noise schedules.
\begin{table}[H]
\centering
\begin{tabular}{|c|c|}
\hline
\textbf{Noise Schedule} & \textbf{Mathematical Expression}  \\
\hline
\text{Linear} & $\beta(t) = \beta_{\min} + (\beta_{\max} - \beta_{\min})t$\\
\hline
\text{Cosine} & $
\beta(t) = \beta_{\min} + (\beta_{\max} - \beta_{\min}) \big(\frac{1}{2} (1 - \cos(\pi t))  \big)$\\
%\hline
%\text{Sigmoid} & $\beta(t) = \beta_{\max} + \frac{\beta_{\min} - \beta_{\max}}{1 + e^{-k(t - 0.5)}}$   \\
%\hline
%\text{Exponential} & $\beta(t) = \beta_{\min} \cdot (\frac{\beta_{\max}}{\beta_{\min}})^{t}$   \\
\hline
\end{tabular}
\caption{Noise schedule}
\label{tab:NS}
\end{table}
\noindent{In Table 1, hyperparameters $(\beta_{\min}, \beta_{\max}) = (0.1, 20)$}, remaining consistent with the settings in \cite{Jonathan-2020, Gabriel-GFDM-2024}.

\medskip

\noindent\textbf{Architecture details for the neural networks} All experiments employed the U-Net \cite{Olaf-2015} architecture to evaluate the designed score function and an Adam optimizer using PyTorch's OneCycle learning rate scheduler. The U-Net architecture is structured with multiple stages, each containing three residual blocks. The architecture applies a channel multiplication strategy defined by the sequence $[1, 2, 2, 2, 2]$, indicating that the number of channels increases progressively from the first stage, and the following stages double the number of feature channels, enhancing the ability of the network to capture increasingly complex and abstract features at deeper layers. The models were trained with a maximal learning rate of $10^{-4}$ for 50k iterations and a batch size of 1,024. The training was conducted on a single Nvidia A6000 GPU, requiring approximately 17~h per training session. Table \ref{tab:hyperparams} summarizes the training hyperparameters.

\begin{table}[H]
    \centering
    \begin{tabular}{ll}
        \toprule
        \textbf{Hyperparameter} & \textbf{Value} \\
        \midrule
        Model architecture& Conditional U-Net \\
        Optimizer & Adam \\
        Learning rate& $10^{-4}$ \\
        Batch size& 1024 \\
        Training steps& 50,000 \\
        Residual blocks& 3 \\
        Channel multiplication& $[1, 2, 2, 2, 2]$ \\
        Input size& $32 \times 32$ \\
        Attention resolution& $[4, 2]$ \\
        Dropout & 0.1 \\
        \bottomrule
    \end{tabular}
    \caption{Training hyperparameter summary.}
    \label{tab:hyperparams}
\end{table}

\medskip

\noindent\textbf{Evaluation metrics} The Fr\'{e}chet inception distance (FID) is a widely used metric in generative modeling to evaluate the quality of generated samples. The FID quantifies the similarity between the distribution of the generated data and that of the real data by computing the Fr\'{e}chet distance between their feature representations. The mathematical formula for the FID is provided in Eq.~\eqref{FID}:
\begin{equation}
\label{FID}
\text{FID} := \| \mu_r - \mu_g \|_2^2 + \operatorname{Tr}\left( \Sigma_r + \Sigma_g - 2(\Sigma_r \Sigma_g)^{\frac{1}{2}} \right),
\end{equation}
where $\mu_g$ and $\Sigma_g$ represent the mean and covariance of the generated distribution, and $\mu_r$ and $\Sigma_r$ represent the mean and covariance of the distribution known to the model. 

\medskip

A lower FID score signifies a higher similarity between the generated and real data, indicating better generative quality. The FID provides a useful and widely adopted measure of generative model quality by capturing differences in distributional statistics. However, as a single scalar value, this metric condenses the comparison between two distributions, making it less informative regarding distinct aspects, such as fidelity (realism of generated samples) and diversity (variety of generated samples) \cite{Mehdi-2018}. Although the FID remains a valuable benchmark metric, it is complementary to other evaluation metrics because it does not explicitly identify one-to-one matches between distributions, can be sensitive to outliers, and depends on specific evaluation settings \cite{Muhammad-2020}. Combining the FID with additional measures provides a more comprehensive assessment of generative model performance. The {\em precision and recall metric} \cite{Mehdi-2018} and its improved version \cite{Karras-2019} were introduced as measures of fidelity and diversity. Nevertheless, the authors in \cite{Muhammad-2020} noted that these two evaluation techniques are unsuitable for practical application because they fail to fulfill the essential criteria for effective evaluation metrics. 

\medskip

To overcome the shortcomings of the FID in distinguishing between distinct aspects of generative quality, fidelity, and diversity, the authors in \cite{Mehdi-2018} presented the {\em density and coverage metrics}, specifically designed to provide a more detailed assessment of generative performance. By incorporating a simple yet carefully designed manifold estimation procedure, this approach enhances the empirical reliability of fidelity-diversity metrics and provides a foundation for theoretical analysis. Table \ref{tab:metrics} presents a concise overview of the listed metrics, whereas the full details of the density and coverage metric are omitted for brevity. 

We assume that it is possible to draw samples $\{X_i\}$ from a real distribution $P(X)$ and $\{Y_j\}$ from a generative model $Q(Y)$, respectively. Let $B(x, r)$ be the sphere in $\mathbb{R}^D$ around $x$ with radius $r$ and let $N$ and $M$ be the number of real and fake samples. The manifold $\mathcal{M}(X_1, \dots, X_N) := \bigcup_{i=1}^N B(X_i, \text{NND}_k(X_i))$, where $\text{NND}_k(X_i)$ denotes the distance from $X_i$ to the $k$th nearest neighbor among $\{X_i\}$, excluding itself. 
\begin{table}[H]
\centering
\begin{tabular}{|c|c|p{5cm}|}
\hline
\textbf{Metric} & \textbf{Mathematical expression}& \textbf{Explanation}  \\
\hline
\textbf{Precision} \cite{Karras-2019} & 
$\displaystyle\frac{1}{M} \sum_{j=1}^M 1_{Y_j \in \mathcal{M}(X_1, \dots, X_N)}$ & 
Measures the proportion of generated samples \( Y_j \) falling in the real data manifold \( \mathcal{M}(X_1, \dots, X_N) \). Higher precision means generated samples are realistic but do not guarantee diversity.\\
\hline
\textbf{Recall} \cite{Karras-2019} & 
$\displaystyle\frac{1}{N} \sum_{i=1}^N 1_{X_i \in \mathcal{M}(Y_1, \dots, Y_M)}$ & 
Evaluates how well the generator covers the real data distribution by checking whether real samples \( X_i \) lie in the generated data manifold \( \mathcal{M}(Y_1, \dots, Y_M) \). Higher recall implies better diversity but does not ensure realism.\\
\hline
\textbf{Density} \cite{Muhammad-2020} & 
$\displaystyle\frac{1}{kM} \sum_{j=1}^M \sum_{i=1}^N 1_{Y_j \in B(X_i, \text{NND}_k(X_i))}$ & 
Quantifies how densely generated samples \( Y_j \) populate the real data space by counting how many real-sample neighborhoods contain \( Y_j \). Higher density indicates greater sample concentration. \\
\hline
\textbf{Coverage} \cite{Muhammad-2020} & 
$\displaystyle\frac{1}{N} \sum_{i=1}^N 1_{\exists j \text{ s.t. } Y_j \in B(X_i, \text{NND}_k(X_i))}$ & 
Measures the proportion of real samples \( X_i \) that have at least one generated sample \( Y_j \) in their neighborhood. Unlike recall, coverage is less sensitive to outliers, offering a more robust assessment of generative diversity. \\
\hline
\end{tabular}
\caption{Comparison of generative model evaluation metrics.}
\label{tab:metrics}
\end{table}

\noindent When evaluating generative models, the range of values that density and coverage can take must be considered. The density is not upper bound by 1 because it considers multiple generated samples per the real data point. A higher density indicates a greater concentration of generated samples. Coverage is not normalized, but higher values are better, meaning the generated data distribution better represents the actual data. Coverage is bounded between $0$ and $1$,  and the maximum value of $1$ indicates that every real sample has at least one generated counterpart in its neighborhood. (See \cite{Karras-2019, Muhammad-2020} and the references therein for a comprehensive understanding of these metrics, including their mathematical definitions and implications.)

\medskip

\noindent\textbf{Experimental results} We downloaded the corresponding protein chain from \url{http://www.pdb.org} and extracted the 3D coordinate information of $\alpha$-carbon ($C_{\alpha}$) atoms for each protein. The number of amino acids considered in the distance matrix is a critical parameter in experiments and influences the resolution and effectiveness of structural modeling. For computational efficiency, we consider only the first $32$ amino acids. The dataset contains $60,000$ proteins and is divided into training ($80\%$) and testing ($20\%$) sets. During each experiment, 12,000 samples were sampled to compute the metrics with the testing set. The models were evaluated based on {\em density and coverage} and the {\em FID}, indicated with arrows $\uparrow$ implying that higher values are better. The best results arein bold. 

\medskip

\noindent\textbf{Importance of modeling} This section presents quantitative comparisons of model performance under various Hurst parameters $H = (0.2, 0.5, 0.8)$ and noise schedules (linear and cosine). 

\medskip

First, we evaluated the generative performance of the FVP model with various selections of $K$ (the number of OU processes to approximate fBm) under a linear noise schedule and compared it with the standard VP-SDE baseline. For all models, we employed the classical Euler--Maruyama scheme to solve the corresponding reverse-time SDE numerically, ensuring consistency across experiments. Table~\ref{tab:comparison1-linear} summarizes the model performance across configurations.
\begin{table}[H]
\centering
\begin{adjustbox}{width=1.0\textwidth} %width=0.9\textwidth for smaller font. 
\begin{tabular}{ccccccccccc}
\toprule
& \multicolumn{3}{c}{$H = 0.2$} & \multicolumn{3}{c}{$H = 0.5$} & \multicolumn{3}{c}{$H = 0.8$} \\
\cmidrule(lr){2-4} \cmidrule(lr){5-7} \cmidrule(lr){8-10}
FVP ({\bf Linear}) & Density $\uparrow$ & Coverage $\uparrow$ & FID $\downarrow$ & Density $\uparrow$ & Coverage $\uparrow$ & FID $\downarrow$ & Density $\uparrow$ & Coverage $\uparrow$ & FID $\downarrow$ \\
\midrule
VP (baseline) & - & - & - & 1.043 & 0.884 & 75.368 &  - &  - & -\\
$K = 2$ & 0.942 & 0.858 & 77.219 & 0.965 & 0.896 & 75.508 &  1.0142 &  0.922 & 74.899 \\
$K = 3$ & 0.932 & 0.840 & 77.174 & 1.028 &0.894 & 75.934 & \textbf{1.118} &  \textbf{0.934} &  \textbf{74.614} \\
%\midrule
%FVP (Cosine) & Density $\uparrow$ & Coverage $\uparrow$ & FID $\downarrow$ & Density $\uparrow$ & Coverage $\uparrow$ & FID $\downarrow$ & Density $\uparrow$ & Coverage $\uparrow$ & FID $\downarrow$ \\
%\midrule
%$K = 3$ & \textbf{-} & - & - & - &  -&  - & - &  -&  - \\
%\midrule
%FVP (Exponential) & Density $\uparrow$ & Coverage $\uparrow$ & FID $\downarrow$ & Density $\uparrow$ & Coverage $\uparrow$ & FID $\downarrow$ & Density $\uparrow$ & Coverage $\uparrow$ & FID $\downarrow$ \\
%\midrule
%$K = 3$ & \textbf{-} & - & - & - &  -&  - & - &  -&  - \\
\bottomrule
\end{tabular}
\end{adjustbox}
\caption{Quantitative results for fractional variance-preserving dynamics with various $H$ and linear noise schedules.}
\label{tab:comparison1-linear}
\end{table}
\noindent The results reveal that increasing the Hurst parameter $H$ generally improves performance, with higher density and coverage and a lower FID at $H = 0.8$. The best overall model is $H = 0.8$, $K = 3$, achieving the highest density ($1.118$) and coverage ($0.934$) and the lowest FID ($74.614$), indicating superior generative quality. Increasing $K$ slightly enhances coverage and the FID but has a minimal effect on the density. Compared to the VP baseline ($H = 0.5$), the FVP models perform better at $H = 0.8$, highlighting the advantage of fractional modeling in capturing complex data distributions.

\medskip

We generate bar charts for the linear noise schedule, separately visualizing the density, coverage, and FID across Hurst parameter $H$ values. Each chart presents the results for $K =2$ and $K = 3$ to facilitate a comparative analysis of their influence on model performance.
\begin{figure}[H]
    \centering
    \begin{minipage}{0.49\textwidth}
        \centering
        \begin{tikzpicture}
            \begin{axis}[
                ybar,
                bar width=15pt,
                width=\textwidth, height=6cm,
                symbolic x coords={H=0.2, H=0.5, H=0.8},
                xtick=data,
                ymin=0.7, ymax=1.2,
                ylabel={Metric Value},
                title={Performance for \( K=2 \)},
                legend style={at={(0.5,-0.15)}, anchor=north, legend columns=-1},
                nodes near coords,
                enlarge x limits=0.2
            ]
            % K=2 Metrics (Density, Coverage)
            \addplot coordinates {(H=0.2, 0.94198) (H=0.5, 1.04304) (H=0.8, 1.01424)}; % Density
            \addplot coordinates {(H=0.2, 0.8582) (H=0.5, 0.8835) (H=0.8, 0.9217)}; % Coverage
            \legend{Density, Coverage}
            \end{axis}
        \end{tikzpicture}
        \caption{Quantitative results for \( K=2 \)}
    \end{minipage}
    \hfill
    \begin{minipage}{0.49\textwidth}
        \centering
        \begin{tikzpicture}
            \begin{axis}[
                ybar,
                bar width=13.5pt,
                width=\textwidth, height=6cm,
                symbolic x coords={H=0.2, H=0.5, H=0.8},
                xtick=data,
                ymin=0.7, ymax=1.2,
                ylabel={Metric Value},
                title={Performance for \( K=3 \)},
                legend style={at={(0.5,-0.15)}, anchor=north, legend columns=-1},
                nodes near coords,
                enlarge x limits=0.2
            ]
            % K=3 Metrics (Density, Coverage)
            \addplot coordinates {(H=0.2, 0.93154) (H=0.5, 1.02764) (H=0.8, 1.11824)}; % Density
            \addplot coordinates {(H=0.2, 0.8399) (H=0.5, 0.8944) (H=0.8, 0.9343)}; % Coverage
            \legend{Density, Coverage}
            \end{axis}
        \end{tikzpicture}
        \caption{Quantitative results for \( K=3 \)}
    \end{minipage}
\end{figure}
\noindent
Table 5 presents the FID bar plots for the linear case, separated for $K=2$ and $K=3$. Table \ref{tab:comparison1-cos} compares model performance under the cosine noise schedule and Hurst parameters $H = (0.2, 0.5, 0.8)$. 
\begin{figure}[H]
\centering
\includegraphics[height=6cm,width=15cm]{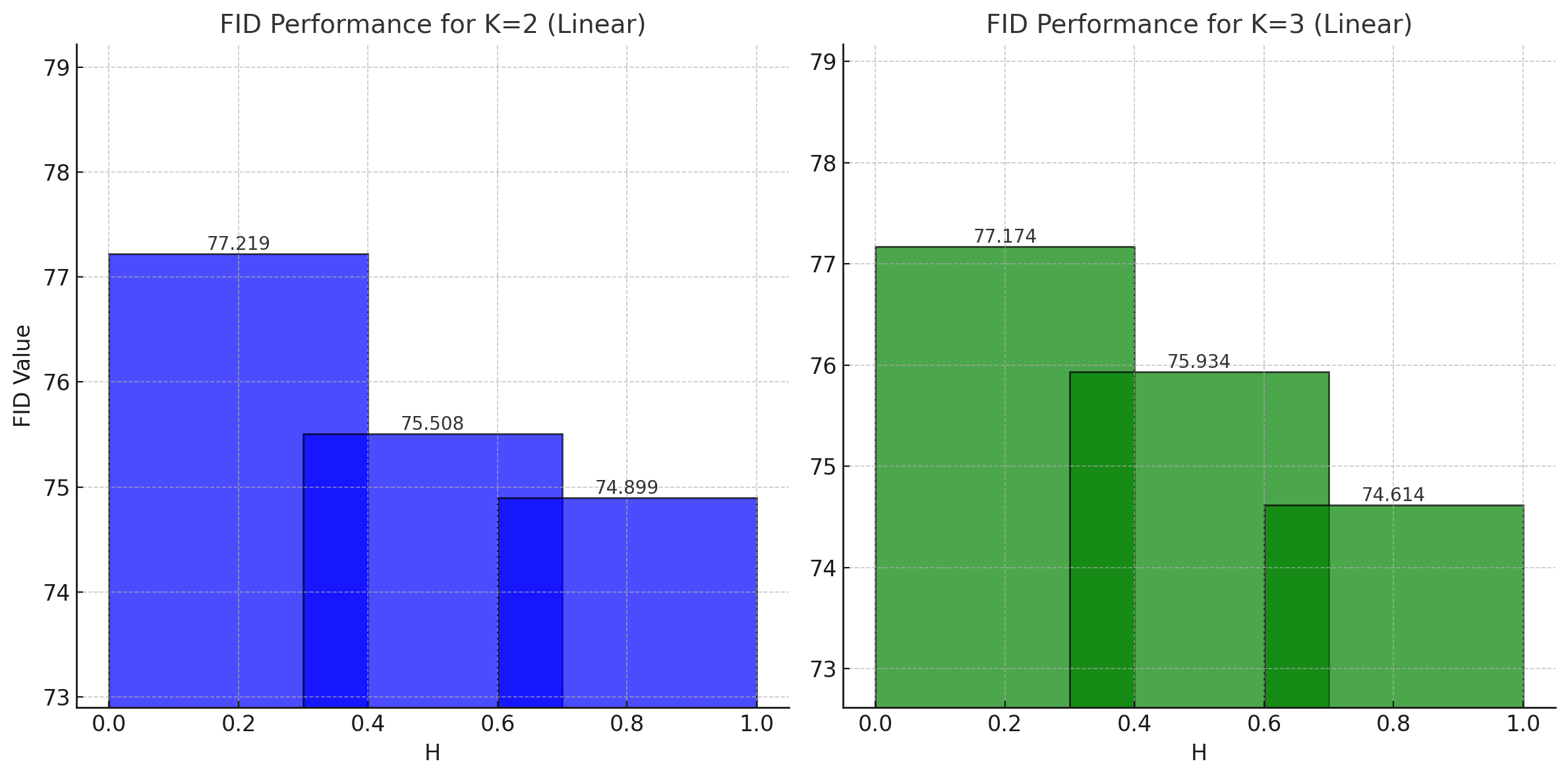}
\caption{Comparison of the Fr\'{e}chet inception distance (FID) performance}
%\label{fig-1}
\end{figure}

\begin{table}[H]
\centering
\begin{adjustbox}{width=1.0\textwidth} %width=0.9\textwidth for smaller font. 
\begin{tabular}{ccccccccccc}
\toprule
& \multicolumn{3}{c}{$H = 0.2$} & \multicolumn{3}{c}{$H = 0.5$} & \multicolumn{3}{c}{$H = 0.8$} \\
\cmidrule(lr){2-4} \cmidrule(lr){5-7} \cmidrule(lr){8-10}
FVP ({\bf Cosine}) & Density $\uparrow$ & Coverage $\uparrow$ & FID $\downarrow$ & Density $\uparrow$ & Coverage $\uparrow$ & FID $\downarrow$ & Density $\uparrow$ & Coverage $\uparrow$ & FID $\downarrow$ \\
\midrule
% VP (baseline) & - & - & - & - & - & - &  - &  - & -\\
$K = 2$  & 0.960 & 0.936 & 74.367 & 0.799 &  0.853 &  78.907 & \textbf{1.04} & \textbf{0.954} & 75.160 \\
$K = 3$ & 0.960 & 0.923 & \textbf{73.452} & 0.832 &  0.861&  79.305 & 1.011 &  0.941&  75.390 \\
%\midrule
%FVP (Cosine) & Density $\uparrow$ & Coverage $\uparrow$ & FID $\downarrow$ & Density $\uparrow$ & Coverage $\uparrow$ & FID $\downarrow$ & Density $\uparrow$ & Coverage $\uparrow$ & FID $\downarrow$ \\
%\midrule
%$K = 3$ & \textbf{-} & - & - & - &  -&  - & - &  -&  - \\
%\midrule
%FVP (Exponential) & Density $\uparrow$ & Coverage $\uparrow$ & FID $\downarrow$ & Density $\uparrow$ & Coverage $\uparrow$ & FID $\downarrow$ & Density $\uparrow$ & Coverage $\uparrow$ & FID $\downarrow$ \\
%\midrule
%$K = 3$ & \textbf{-} & - & - & - &  -&  - & - &  -&  - \\
\bottomrule
\end{tabular}
\end{adjustbox}
\caption{Quantitative results for the fractional variance-preserving dynamics with various $H$ and cosine noise schedules.}
\label{tab:comparison1-cos}
\end{table}
\noindent The results indicate that higher Hurst parameters $H$ improve diversity, as observed in increased coverage, but do not always yield a lower FID. The best FID is achieved at $H = 0.2$ and $K = 3$ (FID = 73.452), suggesting optimal generative quality under the cosine schedule. Moreover, $H=0.8$ and $K=2$ (density = 1.04, coverage = 0.954) offer the best balance between fidelity and diversity.

\medskip

Figure \ref{line-plot-comp} presents line plots comparing density, coverage, and the FID across Hurst parameters $H$ for linear and cosine noise schedules, with separate curves for $K = 2$ and $K=3$. The objective is to analyze how the choice of noise schedule and order $K$ affects model performance across generative quality metrics.
\begin{figure}[H]
\centering
\includegraphics[width=15cm]{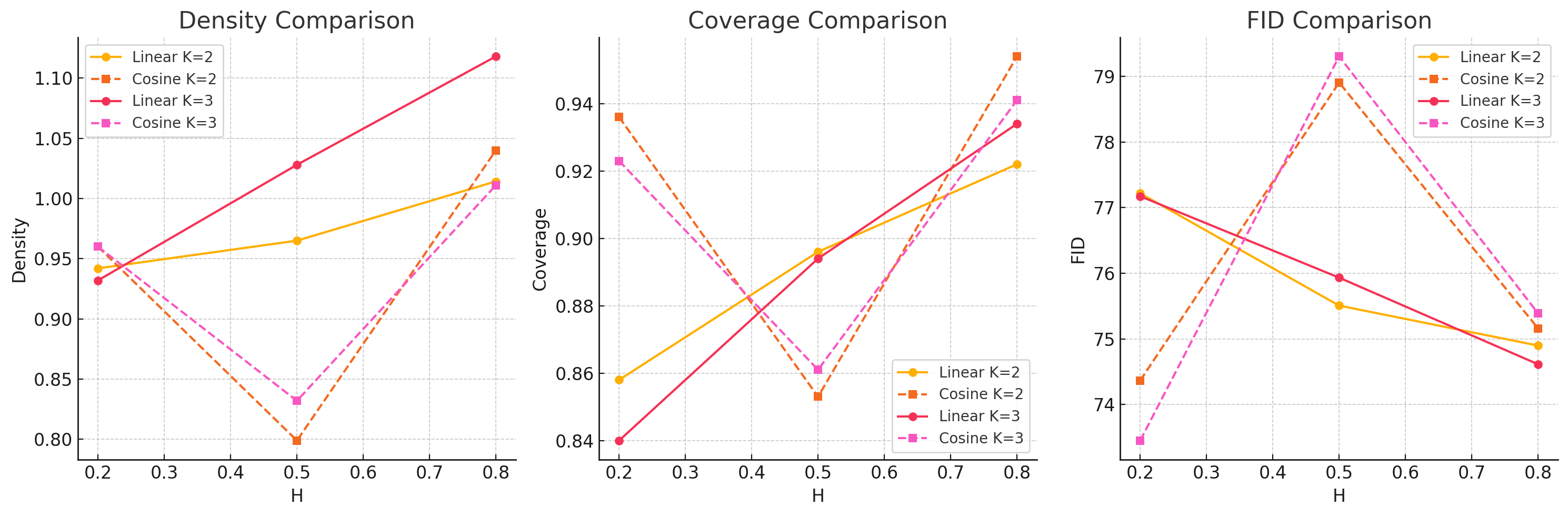}
\caption{Comparison of density, coverage, and the Fr\'{e}chet inception distance (FID) for linear and cosine noise schedules.}
\label{line-plot-comp}
\end{figure}

\noindent Figure \ref{line-plot-comp} reveals that the cosine schedule performs better at lower $H$ values, achieving the best FID at $H=0.2$ and $K=3$, suggesting it preserves fine details early in the diffusion process. In contrast, the linear schedule outperforms others at $H=0.8$, yielding higher density and coverage, indicating better long-range diversity. The FID peaks at $H=0.5 $ for both schedules, implying that BM introduces slight degradation in generative quality. Across all settings, density and coverage increase with $H$, whereas the choice for the noise schedule influences the balance between the sample fidelity and diversity rather than consistently improving performance.

\medskip

\noindent\textbf{Importance of sampling methods} This section evaluates the influence of various SDE and ODE solvers on generative performance. The number of score function evaluations (NFEs) is employed to assess computational efficiency, whereas density, coverage, and the FID measure generative quality. We examined solver performance with 1,000 and 2,000 discretization steps, analyzing how the step count influences sample fidelity and diversity. 

\medskip

First, we considered the SDE solvers, using the classical Euler ODE as a baseline, along with the Euler–Maruyama method and PC sampler. We applied the reverse diffusion SDE as the predictor and Langevin dynamics as the corrector. Table~\ref{tab:SDE_comparison} presents the results.
\begin{table}[H]
\centering
\begin{adjustbox}{width=1.0\textwidth} % Adjust width if needed
\begin{tabular}{cccccccccc}
\toprule
\multicolumn{10}{c}{\textbf{SDE Solver}} \\
\midrule
&  \multicolumn{3}{c}{\text{Euler (baseline)}} & \multicolumn{3}{c}{\textcolor{blue}{Euler-Maruyama}} & \multicolumn{3}{c}{\textcolor{blue}{PC sampler}} \\
\cmidrule(lr){2-4} \cmidrule(lr){5-7} \cmidrule(lr){8-10}
\textbf{Iteration} & Density $\uparrow$ & Coverage $\uparrow$  & FID $\downarrow$ & Density $\uparrow$ & Coverage $\uparrow$  & FID $\downarrow$ & Density $\uparrow$ & Coverage $\uparrow$  & FID $\downarrow$ \\
\midrule
1000 & 0.908  & 0.847 & 74.395 & 1.118 &  0.934 &  74.614  & \textbf{1.224}  & \textbf{0.965}  & \textbf{74.432} \\
2000 & 0.907  & 0.859  & 74.490 & 1.035  & 0.949  & 75.003  & 1.17 & 0.938  & 74.884  \\
\bottomrule
\end{tabular}
\end{adjustbox}
\caption{{\bf Density and coverage} and the {\bf Fr\'{e}chet inception distance (FID)} for  stochastic differential equation (SDE) solvers.}
\label{tab:SDE_comparison}
\end{table}

\noindent The classical Euler ODE baseline provides a reference point but lacks the stochastic flexibility for optimal generative modeling. The PC sampler at 1,000 steps performs best overall, with the highest density (1.224) and coverage (0.965) and the lowest FID (74.432), indicating superior generative quality. Although increasing the discretization steps to 2,000 marginally improves the coverage, it does not consistently enhance the density or FID, particularly for the PC sampler, where a slight performance degradation occurs. The Euler–Maruyama method remains computationally efficient but is outperformed by the PC method, which yields better sample diversity and fidelity at a similar computational cost. Figure~\ref{barplot-20250321-1} displays the comparative bar plots for the density, coverage, and FID metrics across SDE solvers (Euler, Euler–Maruyama, and PC sampler) at 1,000 and 2,000 iterations.
\begin{figure}[H]
\centering
\includegraphics[width=15cm]{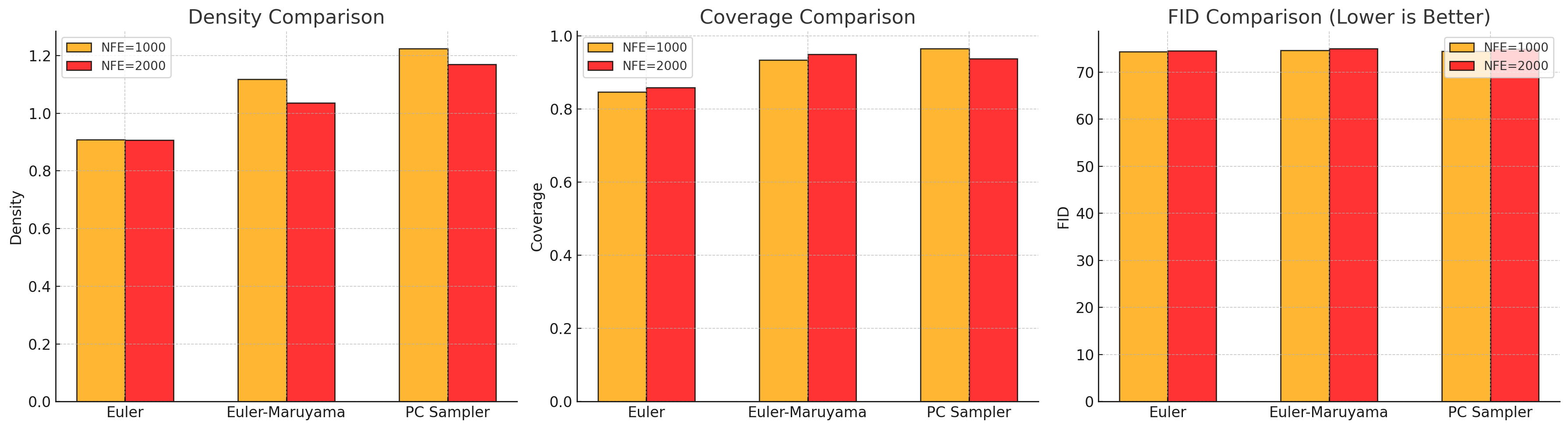}
\caption{Comparison of density, coverage, and Fr\'{e}chet inception distance (FID) metrics for the Euler ordinary differential equation solver and two stochastic differential equation solvers at two iteration counts.}
\label{barplot-20250321-1}
\end{figure}

After examining the influence of various SDE solvers, this work focuses on evaluating the effects of ODE solvers in the generative modeling process. We investigated how several classical ODE solvers influence sample quality, diversity, and computational efficiency. This experiment considers three numerical solvers. The Euler method provides a reference for comparison. The commonly applied fixed-step solver, fourth-order RK (RK4), is known for improved accuracy, and the adaptive-step solver RK 45 (RK45) dynamically adjusts the step size for better precision and efficiency. Table~\ref{tab:Classical_ODE_comparison} summarizes the quantitative results.
\begin{table}[H]
\centering
\begin{adjustbox}{width=1.0\textwidth} % Adjust width if needed
\begin{tabular}{cccccccccc}
\toprule
\multicolumn{10}{c}{\textbf{Classical ODE Solvers}} \\
\midrule
&  \multicolumn{3}{c}{\text{Euler (baseline)}} & \multicolumn{3}{c}{\textcolor{blue}{$4$th order Runge–Kutta (RK4)}} & \multicolumn{3}{c}{\textcolor{blue}{RK45}} \\
\cmidrule(lr){2-4} \cmidrule(lr){5-7} \cmidrule(lr){8-10}
\textbf{Iteration} & Density $\uparrow$ & Coverage $\uparrow$  & FID $\downarrow$ & Density $\uparrow$ & Coverage $\uparrow$  & FID $\downarrow$ & Density $\uparrow$ & Coverage $\uparrow$  & FID $\downarrow$ \\
\midrule
1000 & 0.908  & 0.847 & 74.395 & 0.83 &  0.75 &  77.9  & \textbf{1.00}  & \textbf{0.96}  & \textbf{73.8} \\
2000 & 0.907  & 0.859  & 74.490 & 0.886  & 0.862  & 74.3  & 0.973 & 0.89  & 74.2  \\
\bottomrule
\end{tabular}
\end{adjustbox}
\caption{{\bf Density and coverage} and the {\bf Fr\'{e}chet inception distance} for ordinary differential equation solvers.}
\label{tab:Classical_ODE_comparison}
\end{table}
\noindent Table \ref{tab:Classical_ODE_comparison} reveals that the RK45 solver at 1,000 steps performs best, achieving the highest density (1.00) and coverage (0.96) and the lowest FID (73.8), indicating superior sample fidelity and diversity. Although RK4 displays degraded FID performance, the Euler method remains a computationally simple alternative but is outperformed by RK45. Increasing discretization steps marginally improves coverage but does not significantly enhance the FID, suggesting that adaptive solvers, such as RK45, balance efficiency and accuracy better than fixed-step methods. The bar plots in Figure~\ref{barplot-20250321-2} compare the density, coverage, and FID metrics for ODE solvers (Euler, RK4, and RK45) evaluated at 1,000 and 2,000 iterations.
\begin{figure}[H]
\centering
\includegraphics[width=15cm]{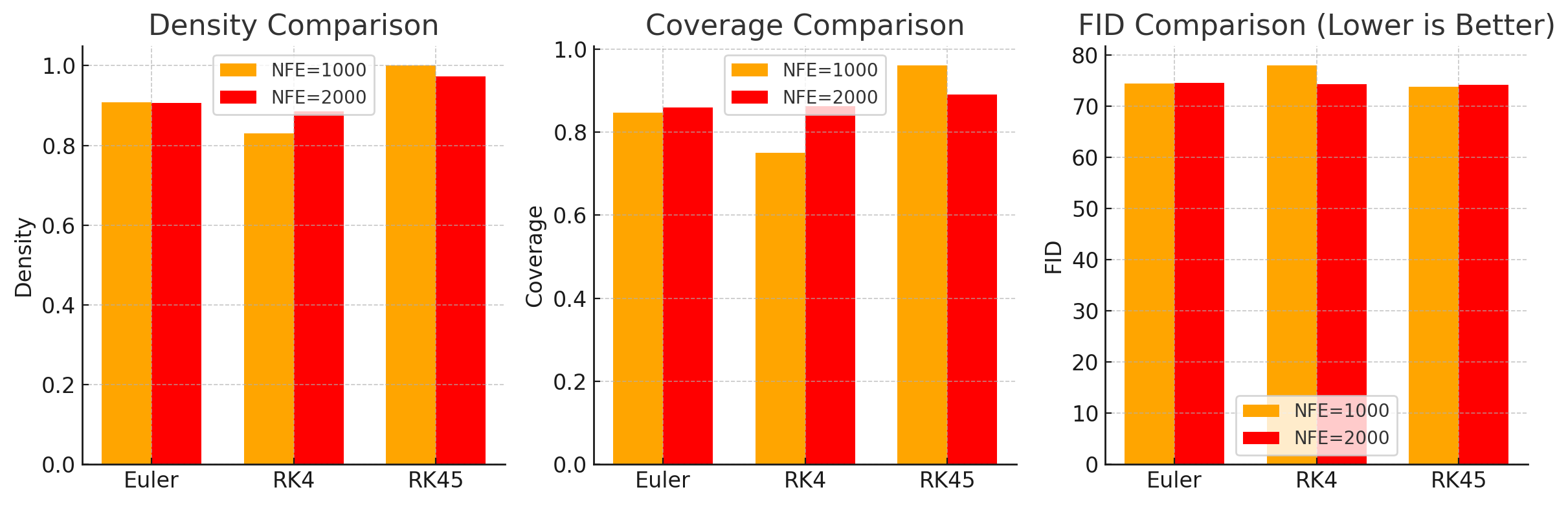}
\caption{Comparison of density, coverage, and the Fr\'{e}chet inception distance (FID) metrics for ordinary differential equation solvers at two iteration counts.}
\label{barplot-20250321-2}
\end{figure}
% \noindent We observe that the RK45 method achieves the best overall performance, with the highest Density and Coverage, and the lowest FID, making it the most effective solver among the three.
% \begin{table}[H]
% \centering
% \begin{adjustbox}{width=0.8\textwidth} % Adjust width if needed
% \begin{tabular}{ccccccc}
% \toprule
% \multicolumn{7}{c}{\textbf{Fast Sampling ODE Solver}} \\
% \midrule
% & \multicolumn{3}{c}{\text{DPM Solver-2}} & \multicolumn{3}{c}{\text{$t$AB}} \\
% \cmidrule(lr){2-4} \cmidrule(lr){5-7}
% \textbf{NFE} & Density $\uparrow$ & Coverage $\uparrow$  & FID $\downarrow$ & Density $\uparrow$ & Coverage $\uparrow$  & FID $\downarrow$ \\
% \midrule
% 50 & - & - & - & - & - & -  \\
% 100 & - & - & - & - & - & -  \\
% 250 & - & - & - & - & - & -  \\
% 500 & - & - & - & - & - & -  \\
% \bottomrule
% \end{tabular}
% \end{adjustbox}
% \caption{{\bf Density \& Coverage} and {\bf FID} for various ODE solvers}
% \label{tab:ODE_comparison}
% \end{table}
%\cleardoublepage
%%%%%%%%%%%%%%%%%%%%%%%%%%%%%%%%%%%%%%%%%%%%%

%%%%%%%%%%%%%%%%%%%%%%%%%%%%%%%%%%%%%%%%%%%%
\section{Conclusion and Future work}
\label{Conclusion}
{\bf Conclusion} This work proposes ProT-GFDM, a novel generative framework for protein generation. This model employs fractional stochastic dynamics to model long-range dependencies in protein backbone structures more effectively. The detailed analysis highlights the importance of modeling choices, demonstrating that the Hurst parameter, noise scheduling, and solver selection factors significantly influence generative performance. 

\medskip

The quantitative results in Tables~\ref{tab:comparison1-linear} and \ref{tab:comparison1-cos} highlight the Hurst parameter $H$ and the number of OU processes $K$ as critical modeling factors that significantly influence generative performance. Based on the results in those tables, higher $H$ generally improves density and coverage in both noise schedules, suggesting that higher Hurst parameters allow the model to capture long-range dependencies better. The FID trends vary, with a lower FID observed at $H=0.2$ for the cosine schedule (best FID = $73.452$, $K=3$), whereas, at $H=0.8$, the linear schedule achieves a lower FID (best FID = $74.614$, $K=3$), implying that noise scheduling influences generative quality differently across Hurst values. Increasing $K$ from $2$ to $3$ results in marginal improvements in coverage and the FID but does not significantly affect density, indicating that additional OU processes may slightly refine the sample diversity but do not drastically alter generative performance.

\medskip

Most previous literature studies have focused on experiments with linear noise schedules, whereas this work explores an alternative nonlinear noise schedule, the cosine noise schedule, to demonstrate how the choice of noise schedule affects model performance. 
The cosine noise schedule is introduced to explore its potential practical benefits, particularly regarding performance enhancement and stability. However, as indicated by the comparison between linear (Table~\ref{tab:comparison1-linear}) and cosine (Table~\ref{tab:comparison1-cos}) noise schedules, changing the noise schedule does not consistently improve performance across all metrics . Still, it does influence the trade-off between fidelity and diversity. The cosine schedule achieves the lowest FID at $H=0.2$ and $K=3$ (FID = 73.452), suggesting better generative quality in some cases. In contrast, the linear schedule yields higher density and Coverage at $H=0.8,K=3$ (Density = 1.118, Coverage = 0.934), favoring sample diversity. The cosine schedule tends to preserve more structure in early diffusion steps but then applies noise more aggressively in later stages. This may lead to suboptimal score-matching behavior, where the model struggles to accurately learn denoising at high noise levels. Some samplers may be better suited for linear noise decay, while cosine-based schedules may require a more specialized adaptive step-size or fine-tuned solver. The interaction between the choice of noise schedule and the sampler remains unclear. There might be other potential factors that could explain this outcome. A very recent paper \cite{Sun2025-noise} challenges the prevailing assumption that noise conditioning is essential for the success of denoising diffusion models. Their findings reveal that most models degrade gracefully without noise conditioning, and some even perform better. This study encourages the research community to reconsider the foundational assumptions and formulations of denoising generative models. 

\medskip

Our findings from Table \ref{tab:SDE_comparison} and Table \ref{tab:Classical_ODE_comparison} underscore the significant impact of sampling methods on generative performance. Comparing SDE and ODE solvers, we observe that the Predictor-Corrector (PC) Sampler outperforms Euler-based methods, achieving the highest Density ($1.224$), Coverage ($0.965$), and lowest FID ($74.432$) at $1000$ iterations. This suggests that correcting steps via Langevin dynamics enhances sample fidelity and diversity. For ODE solvers, the adaptive RK45 solver demonstrates superior performance, reaching the best FID ($73.8$) at $1000$ iterations, outperforming Euler and RK4. While increasing the number of iterations ($2000$ steps) marginally improves Coverage, it does not consistently enhance FID, indicating that an increase in the discretization steps does not always improve performance, with PC Sampler and RK45 slightly degrading at $2000$ steps, possibly due to overcorrection effects. In practice, selecting an appropriate solver for the ScoreSDE model requires balancing computational cost (measured by the NFEs) with sample quality, as reflected by density, coverage, and FID. However, the theoretical understanding of the qualitative differences between sampling from the SDE and the ODE remains an open question, highlighting the need for further analysis.

%the PC Sampler achieves the highest Density (1.224) and Coverage (0.965) at 1000 iterations, along with a low FID (74.432). 
\medskip

\noindent\textbf{Future work} For the driving process, the choice of the number of OU processes, $K$, used to approximate the fBm depends on the data. There is a trade-off: a higher $K$ provides a more accurate approximation of fBm but comes at a higher computational cost. Regarding the Hurst index, we use the classical case \( H = \frac{1}{2} \), while for super-diffusion ($H > \frac{1}{2}$), we set $ H = 0.8$, and for sub-diffusion ($H < \frac{1}{2}$), we set $H = 0.2$. The selection of the Hurst index in these experiments is based on practical considerations, as the true value is unknown. A more precise estimation of the Hurst index could be directly derived from the data \cite{Rembert-2024}.

\medskip

More advanced ODE solvers can also be considered. Recall the general form of Diffusion ODE
\begin{equation}
\label{eqn1234}
d \bold{x}_t  = \big\{ f(t) \bold{x}_t - \frac{1}{2} g^2(t) \bold{s}_{\theta}(\bold{x}_t, t) \big\} dt.
\end{equation}
The model is characterized by a {\bf semi-linear structure}, comprising two main components: a linear function of the data variable $\bold{x}_t$ and a nonlinear function of $\bold{x}_t$ parameterized by neural networks $s_{\theta}(\hat{\bold{x}}, t)$. The solution at time t can be exactly formulated by the ``variation of constants" formula:
$$
\mathbf{x}_t = e^{\int_s^t f(\tau) \, d\tau} \mathbf{x}_s + \int_s^t \left( e^{\int_\tau^t f(r) \, dr} \frac{1}{2} g^2(\tau) \bold{s}_{\theta}(\mathbf{x}_\tau, \tau) \right) d\tau.
$$
We can consider DPM-Solver \cite{Cheng-2022} and DEIS (Diffusion Exponential Integrator Sampler) \cite{Qinsheng-2022}, which aim to solve the PF-ODE in diffusion models efficiently. They share a key similarity: both exploit the semi-linear structure of the PF-ODE to improve efficiency and accuracy. Lu et al. \cite{DPM++} introduced an improved version of the DPM solver with {\em DPM-solver++}, a high-order solver for the guided sampling of DPMs, to further speed up high-quality sample generation. For DEIS, the authors of \cite{DEIS-SN} proposed a modified version of DEIS, called the {\em DEIS-SN}, based on a simple new score parametrization – to normalize the score estimate using its average empirical absolute value at each timestep (computed from high NFE offline generations). This leads to consistent improvements in FID compared to classical DEIS. 

\medskip

\noindent\textbf{Acknowledgements} The authors gratefully acknowledge the support of the Artificial Intelligence for Design Challenge program from the National Research Council Canada for funding this project.

%\cleardoublepage
%%%%%%%%%%%%%%%%%%%%%%%%%%%%%%%%%%%%%%%%%%%%%

%%%%%%%%%%%%%%%%%%%%%%%%%%%%%%%%%%%%%%%%%%%%
\section{Notational conventions}
\label{Notational-conventions}
Definitions of the mathematical symbols used in this work.
\begin{table}[H]
\centering
\resizebox{\textwidth}{!}{
\small
\begin{tabular}{ll}
\hline
\textbf{Symbol} & \textbf{Definition} \\
\hline
$[0, T]$ & Time horizon with terminal time $T > 0$ \\
$X = (X_t)_{t \in [0, T]}$ & Stochastic forward process taking values in $\mathbb{R}$ \\
$D \in \mathbb{N}$ & Data dimension \\
$\mathbf{X}$ & Vector-valued stochastic forward process $\mathbf{X} = (X_t)_{t \in [0, T]}$, $X_t = (X_{t,1}, \ldots, X_{t,D})$ \\
%$\overline{\mathbf{X}}$ & Reverse-time stochastic process with $\overline{X}_t = X_{T - t}$ \\
$f$ & Function $f : \mathbb{R}^D \times [0, T] \rightarrow \mathbb{R}^D$ \\
$\mu, g$ & Functions $\mu, g : [0, T] \rightarrow \mathbb{R}$ \\
%$\overline{f}$ & Reverse-time function with $\overline{f}(x, t) = f(x, T - t)$ \\
%$\overline{\mu}, \overline{g}$ & Reverse-time functions with $\overline{\mu}(t) = \mu(T - t)$ and $\overline{g}(t) = g(T - t)$ \\
$p_0$ & Data distribution \\
$p_t$ & Marginal density of (augmented) forward process at $t \in [0, T]$ \\
$B$ & Brownian motion (BM) \\
$H$ & Hurst index $H \in (0, 1)$ \\
$W^H$ & Type I fractional Brownian motion (fBm) \\
$B^H$ & Type II fractional Brownian motion (fBm) \\
$Y^\gamma = (Y^\gamma_t)_{t \in [0, T]}$ & Ornstein–Uhlenbeck (OU) process with speed of mean reversion $\gamma \in \mathbb{R}$ \\
$K \in \mathbb{N}$ & Number of approximating processes \\
$\gamma_1, \ldots, \gamma_K$ & Geometrically spaced grid \\
$\omega_1, \ldots, \omega_K$ & Approximation coefficients \\
$\omega$ & Optimal approximation coefficients $\omega = (\omega_1, \ldots, \omega_K)$ \\
%$\tilde{\omega}$ & Sum of optimal approximation coefficients \\
$\tilde{B}^H$ & Markov-approximate fractional Brownian motion (MA-fBm) \\
$k$ & $k \in \mathbb{N}$ with $1 \leq k \leq K$ \\
$Y^k$ & OU processes $Y^k = Y^{\gamma^k}$ \\
$Y^{1}, \ldots, Y^K$ & Augmenting processes with $Y^k = (Y^k, \ldots, Y^K)$ \\
$\mathbf{F}, \mathbf{G}$ & Vector-valued functions $\mathbf{F}, \mathbf{G} : [0, T] \rightarrow \mathbb{R}^{D \cdot (K + 1)}$ \\
%$\overline{\mathbf{F}}, \overline{\mathbf{G}}$ & Reverse-time vector-valued functions with %$\overline{\mathbf{F}}(t) = \mathbf{F}(T - t)$ and $\overline{\mathbf{G}}(t) = \mathbf{G}(T - t)$ \\
$\mathbf{Z}$ & By $Y^1, \ldots, Y^K$ augmented forward process \\
$Y^{[K]}$ & Stacked vector of augmenting processes \\
$q_t$ & Marginal density of $Y^{[K]}$ at $t \in [0, T]$ \\
$\theta$ & Weight vector of a neural network \\
\hline
\end{tabular}
}
\end{table}

\cleardoublepage
%%%%%%%%%%%%%%%%%%%%%%%%%%%%%%%%%%%%%%%%%%%%%

%%%%%%%%%%%%%%%%%%%%%%%%%%%%%%%%%%%%%%%%%%%%

%%%%%%%%%%%%%%%%%%%%%%%%%%%%%%%%%%%%%%%%%%%%%

\end{document}